\documentclass[preprint]{aastex6}
\usepackage[utf8]{inputenc}
\usepackage{natbib}
\usepackage{wasysym}
\usepackage{afterpage}
\usepackage{placeins}
\usepackage{float}

\begin{document}

\title{VARIABILITY OF DISK EMISSION IN PRE-MAIN SEQUENCE AND RELATED STARS. IV. INVESTIGATING THE STRUCTURAL CHANGES IN THE INNER DISK REGION OF MWC 480}

\author{Rachel B. Fernandes\altaffilmark{1,2}, Zachary C. Long\altaffilmark{1}, Monika Pikhartova\altaffilmark{1}, Michael L. Sitko\altaffilmark{1,3}, Carol A. Grady\altaffilmark{4}, Ray W. Russell\altaffilmark{5} David M. Luria\altaffilmark{1}, Dakotah B. Tyler\altaffilmark{1}, Ammar Bayyari\altaffilmark{1},William Danchi\altaffilmark{6}, John P. Wisniewski\altaffilmark{7}}

\altaffiltext{1}{Department of Physics, University of Cincinnati, Cincinnati, OH 45221, USA}
\altaffiltext{2}{University of Arizona Lunar and Planetary Laboratory, 1629 E University Blvd, Tucson, AZ 85721, USA}
\altaffiltext{3}{Center for Extrasolar Planetary Systems, Space Science Institute,  4750 Walnut St, Suite 205, Boulder, CO 80301, USA}
\altaffiltext{4}{Eureka Scientific, 2452 Delmer St. Suite 100, Oakland CA 96402, USA}
\altaffiltext{5}{The Aerospace Corporation, PO Box 92957, M2-266, Los Angeles, CA 90009, USA}
\altaffiltext{6}{NASA Goddard Space Flight Center, 8800 Greenbelt Rd, Greenbelt, MD 20771, USA}
\altaffiltext{7}{Homer L. Dodge Department of Physics and Astronomy, University of Oklahoma, 440 W. Brooks St., Norman, OK 73019, USA}

\begin{abstract}
We present five epochs of near IR observations of the protoplanetary disk around MWC 480 (HD 31648) obtained with the SpeX spectrograph on NASA's Infrared Telescope Facility (IRTF) between 2007 and 2013, inclusive. Using the measured line fluxes in the Pa $\beta$ and Br $\gamma$ lines, we found the mass accretion rates to be (1.43 - 2.61)$\times$10$^{-8}$M$_{\astrosun}$y$^{-1}$ and (1.81 - 2.41)$\times$10$^{-8}$M$_{\astrosun}$y$^{-1}$ respectively, but which varied by more than 50\% from epoch to epoch. The spectral energy distribution (SED) reveals a variability of about 30\% between 1.5 and 10 microns during this same period of time. We investigated the variability using  of the continuum emission of the disk in using the Monte-Carlo Radiative Transfer Code (MCRT) HOCHUNK3D. We find that varying the height of the inner rim successfully produces a change in the NIR flux, but lowers the far IR emission to levels below all measured fluxes. Because the star exhibits bipolar flows, we utilized a structure that simulates an inner disk wind to model the variability in the near IR, without producing flux levels in the far IR that are inconsistent with existing data. For this object, variable near IR emission due to such an outflow is more consistent with the data than changing the scale height of the inner rim of the disk.

\end{abstract}

\keywords{Stars: individual (MWC 480) -- stars: variables (Herbig Ae/Be) -- planetary systems: protoplanetary disks -- planetary systems: planet-disk interactions}

\section{Introduction}

MWC 480 is a Herbig Ae star with a circumstellar disk has been the focus of numerous investigations during the past decade. The thermal emission of the dust disk has been detected at millimeter and sub-millimeter wavelengths \citep{mks97,hughes13,andrews13}. It was also detected in scattered light at 1.6 $\mu$m by \citet{Grady10} using HST (Near Infrared Camera and Multi-Object Spectrometer (NICMOS) coronographic imagery. Outflowing jets of material were seenin the NICMOS images, and at visible wavelengths using the Goddard Fabry-Perot Interferometer at the Apache Point Observatory. They demonstrated that the star drives parsec-scale bipolar jets with condensations or knots. Such jets might be launched near the inner rim of the dust disk, similar to what is suggested for HD 163296 by \citet{Ellerbroek14} and more generally by \citet{Bans12}. 

The SED of MWC 480 is consistent with it being a Meeus Group II objects \citep{me01}, which have disks that were postulated to be significantly shadowed, a likely sign of dust grain growth and settling. As such behavior is a sign of disk evolution and aging, it is surprising that it exhibits many indicators of the active accretion generally associated with flared unshadowed Meeus Group I objects such as hot, non-stellar gas and jets. Both imaging by \citet{Kusakabe12} and interferometry combined with the system's spectral energy distribution (SED) by \citet{Millan-Gabet16} suggest a disk whose ratio of scale height to radial distance is small. The imaging suggested a scale height for the dust at a radial distance of 100 AU of only 30\% that of the CO gas scale height of 10$\pm$1.1 AU deduced from modeling CO interferometric data by \citet{Pietu07}. Interferometric /SED data by also indicated a thinner disk than most of the objects in their sample.\\

The NIR  emission of MWC 480 is known to vary by $\sim$30 percent \citep{Sitko08, Grady10, Kusakabe12}. Such NIR variability is not uncommon in pre-main sequence stars, and it has been suggested that such variability is due to changes in the scale height of the ``puffed up'' inner rim of the dust disk \citep{Muzerolle09, Flaherty10, Espaillat11}. Changing the scale height of this optically thick structure is thought to be the source of the ``see-saw'' variability exhibited by many of these disk systems, with the near IR and mid IR fluxes changing in opposite directions. In this scenario, as the scale height of the inner rim increases, it produces an increase in near-IR flux (as the solid angle subtended by the disk rim ``seen'' by the star increased). This simultaneously decreases the emission at longer wavelengths, as more distant portions of the disk experience increased shadowing by the inner disk wall. It should be noted that this behavior is not universal, however, as not all such objects clearly follow this behavior - only about half of the variable sources observed by \citet{Espaillat11} using the Infrared Spectrograph (IRS) of the \textit{Spitzer Space Telescope} (5-38 $\mu$m) actually exhibit this behavior. But if this paradigm applies to MWC 480, the historically low NIR flux seen in 2010 by \citet{Kusakabe12} implied the smallest inner rim height in many years, minimizing the shadow it casts, which coincided with the first significant detection of the outer disk in scattered light. \\

A variable inner rim scenario is only one way that variability in NIR fluxes might arise. Numerous studies have suggested that much of the near-IR emission seen in young stellar disks might have been caused by a wind launched from the inner region of the disks which include material from beyond the dust sublimation radius. \citet{Bans12} suggested a bi-layer wind that includes a dusty wind originating in the disk beyond the dust sublimation radius, and a dust-free gaseous wind launched closer to the star. Here, the dusty wind produces emission from material at temperatures of 1500 Kelvins (a ``canonical'' sublimation temperature often used for silicate grains) and cooler. Dust at these temperatures tends to radiate at wavelengths longer than $\sim$2 $\mu$m and so, in order to increase the flux needed to match the SEDs in these systems required a hotter component - the gas. Interferometric measurements of some young stellar disks, such as that of HD 163296, require that some material be present closer to the star than the canonical dust sublimation radius, but whether this is gaseous or a population of super-refractory grains, is a matter of debate \citep{Tannirkulam08a,Tannirkulam08b, Benisty10}.\\

A disk wind that is associated with the Herbig-Haro (HH) knots in the observed jets in HD 163296 \citep{wassell06} has been detected in ALMA observations in the CO  gas \citep{klaassen13}. An example of disk wind that included a significant dust component was suggested by \citet{Ellerbroek14} for producing the drops in brightness at visible wavelengths in HD 163296 during the past 2 decades. In HD 163296, a $\sim$1 magnitude drop in the light at visible wavelengths was observed in 2001, about the time that one ejection event was to have occurred. Such material, illuminated by the star, could also produce changes in the near-IR emission also seen within a year of that event, and has been confirmed in a radiative transfer model by \citet{Pikhartova18}. In the case of HD 163296, it was necessary to have a substantial optical depth in the jet to reproduce the observed variability. But the observed presence of a jet in MWC 480 suggested that a similar model for it.\\

In this paper, we investigate whether or not the variable inner-rim scenario is consistent with both the existing observed variable NIR emission, the thinness of the disk, and of other existing mid- and far-IR data. The variable inner-rim scenario predicts that there should be significant time-dependent changes in the far-IR flux.  Although numerous epochs of simultaneous observations covering the relevant wavelength range are lacking, they should record noticeable scatter in their values, measuring higher and lower flux states.  We will also determine whether whether an alternative model, one where changes in the outflowing jet material can produce the observed variable NIR emission, and its effect on the mid- and far-IR flux, provides an equally good or better fit to the observations.\\

\section{Observations}
\subsection{SpeX}

Near-IR spectra of MWC 480 were obtained on 5 nights between 2007 and 2013, using the SpeX spectrograph \citep{rayner03} on NASA's Infrared Telescope Facility. The SpeX observations were made using the cross-dispersed (``XD'') echelle gratings between 0.8-5.4 $\mu$m, using an entrance slit width of 0.8 arcsec.  The spectra were processed using a Spextool reduction package \citep{cushing04} running under IDL. In Spextool, telluric corrections and flux calibrations are performed using A0V stars \citep{vacca03} as ``Vega analogs". For MWC 480, the A0V stars HD 25152, HD 31592, HD 31295, and HD 31039 were used for the calibration.  The observing and data reduction procedures were the same as described by \citet{Sitko12}. Beginning with the December 2007 observations, and continuing through 3 of the remaining 4 epochs, we also obtained spectra using the Prism disperser and a 3.0$\arcsec$ wide slit. This technique provides absolute fluxes to an accuracy of $\sim$5\% when the seeing was 1.0$\arcsec$ or better.  \\

\subsection{BASS Spectrophotometry}

We observed MWC 480 on five epochs between 2007 and 2011 using The Aerospace Corporation's  Broad-band Array Spectrograph System (BASS).  BASS uses a cold beamsplitter to separate the light into two separate wavelength regimes. The short-wavelength beam includes light from 2.9-6 $\mu$m, while the long-wavelength beam covers 6-13.5 $\mu$m. Each beam is dispersed onto a 58-element Blocked Impurity Band (BIB) linear array, thus allowing for simultaneous coverage of the spectrum from 2.9-13.5 $\mu$m. The spectral resolution $R = \lambda$/$\Delta\lambda$ is wavelength-dependent, ranging from about 30 to 125 over each of the two wavelength regions \citep{hackwell90}. The entrance aperture of BASS is a 1-mm circular hole, whose effective projected diameter on the sky was $\sim$3.5 ''.The observations are calibrated against spectral standard stars located close to the same airmass. Due to its proximity in the sky, $\alpha$ Tau usually serves as the flux calibration star. \\

\begin{deluxetable}{lcccc}
\tablecolumns{5}
\tablewidth{0pc}
\tablecaption{SpeX Observations}
\tablehead{
\colhead{UT Date} & \colhead{Mode} & \colhead{MWC 480 Airmass} &  \colhead{Calibration Star Airmass} & \colhead{Calibration Star} }
\startdata
2007.12.09 & SXD & 1.03 & 1.06 & HD 25152 \\  
2007.12.10 & LXD &1.28 & 1.11 & HD 25152 \\
                   & Prism & 1.05 & 1.06 & HD 25152 \\
\\
2008.10.04 & SXD & 1.12 & 1.21 & HD 31069 \\ 
                      & LXD & 1.09 & 1.23 &  HD 31069 \\
                      & Prism & 1.04 & 1.01 &  HD 31069 \\
\\
2009.12.01 & SXD & 1.20 & 1.16 & HD 31069 \\
                      & LXD & 1.16 & 1.19 &  HD 31069 \\
                      & Prism & 1.70 & 1.66 & HD 31069 \\
\\
2011.10.016 & SXD &1.02 & 1.07 & HD 25152 \\ 
                     & LXD & 1.02 & 1.14 & HD 25152 \\
                      & Prism & 1.02 & 1.05 & HD 25152 \\
\\
2013.09.11 & SXD & 1.18 & 1.23 & HD 31069 \\
                   & LXD & 1.10 & 1.15 & HD 31069 \\ 
                   & Prism & 1.05 & 1.13 & HD 31069 \\
\enddata
\end{deluxetable}

\begin{deluxetable}{lccc}
\tablecolumns{4}
\tablewidth{0pc}
\tablecaption{BASS Observations}
\tablehead{
\colhead{UT Date}  & \colhead{MWC 480 Airmass} &  \colhead{Calibration Star Airmass} & \colhead{Calibration Star}}
\startdata
1996.10.14 & 1.51-1.66  & 1.59-1.61 & $\alpha$ Tau \\
2004.08.05 & 1.28-1.32  & 1.17-1.18 & $\alpha$ Tau \\
2006.12.11 & 1.08-1.10  & 1.06-1.07 & $\alpha$ Tau \\
2007.08.20 & 1.13-1.15  & 1.13-1.14 & $\alpha$ Tau \\
2008.09.03 & 1.35-1.39  & 1.33-1.35 & $\alpha$ Tau  \\
2009.11.29 & 1.49-1.56  & 1.56-1.60 & $\alpha$ Tau \\
2010.10.23 & 1.05-1.06  & 1.05-1.06 & $\alpha$ Tau \\
2011.10.15 & 1.03-1.04 & 1.03-1.04 & $\alpha$ Tau \\
\enddata
\end{deluxetable}

\subsection{Archival Data}
In order to construct a complete spectral energy distribution for constraining models of the MWC 480 system, we incorporate a variety of archival data, both via the Simbad/Vizier SED tool and other sources in the literature (the Spitzer Heritage Archive, Mikulski Archive for Space Telescopes, Infrared Space Observatory, various optical-infrared photometric surveys \citep{Oudmaijer01, Eiroa01, dewinter01}, 2MASS, AKARI, Herschel PACS, Hubble NICMOS \citep{Grady10}, millimeter surveys \citep{Pietu06, ms97},  the University of Wisconsin's Low Resolution Spectrophotometer (0.34-0.58 $\mu$m), the Kitt Peak National Observatory infrared photometers \citep{Sitko81}, and the HiCIAO photometric point from  \citet{Kusakabe12} . These data were obtained over a span of many decades and are not contemporaneous with the SpeX observations. Nevertheless, some are uniquely suited for placing useful limits on the nature of the SED from ultraviolet to millimeter wavelengths., in particular, if there is evidence of scatter due to variability at mid-IR wavelength, such as those expected for ``see-saw'' behavior predicted by the variable inner disk rim model.\\

\subsection{Near-IR Variability of MWC 480}
Figure 1 shows the new and archival flux data on MWC 480. In most cases where both SpeX and BASS data were obtained within a few days to a few weeks of one another, the two independent data sets appear to be in agreement with one another, although variability on time scales of days cannot be entirely ruled out, without a more extensive set of observations with a short (days) cadence. \\

We also show the difference in the ``high state'' and ``low state'' of the SpeX-derived fluxes. It clearly shows that the bulk of the difference is consistent with a change in emission of a component whose temperature is $\sim$ 1600 K, indicating that the changes are occurring close to the sublimation temperature of silicate grains, and hence close to the inner edge of the dust disk. Additional flux at shorter wavelengths is likely coming from gas emission, much of which may occur inside the dust sublimation radius. \\

Some of the flux levels we observed in the near-IR would seem to be lower than the bulk of historic data obtained from archival data. It has been suggested that the HiCIAO detection of the disk in scattered light in 2010, compared to more marginal detections at other epochs, might have been due to a minimum in the shadowing of the outer disk by the inner disk wall, often assumed to have a slightly ``puffed-up'' geometry \citep{Kusakabe12}. However, we will show that another scenario may be implicated in the flux variability of MWC 480. \\

\begin{figure}[H]
\figurenum{1}
\plotone{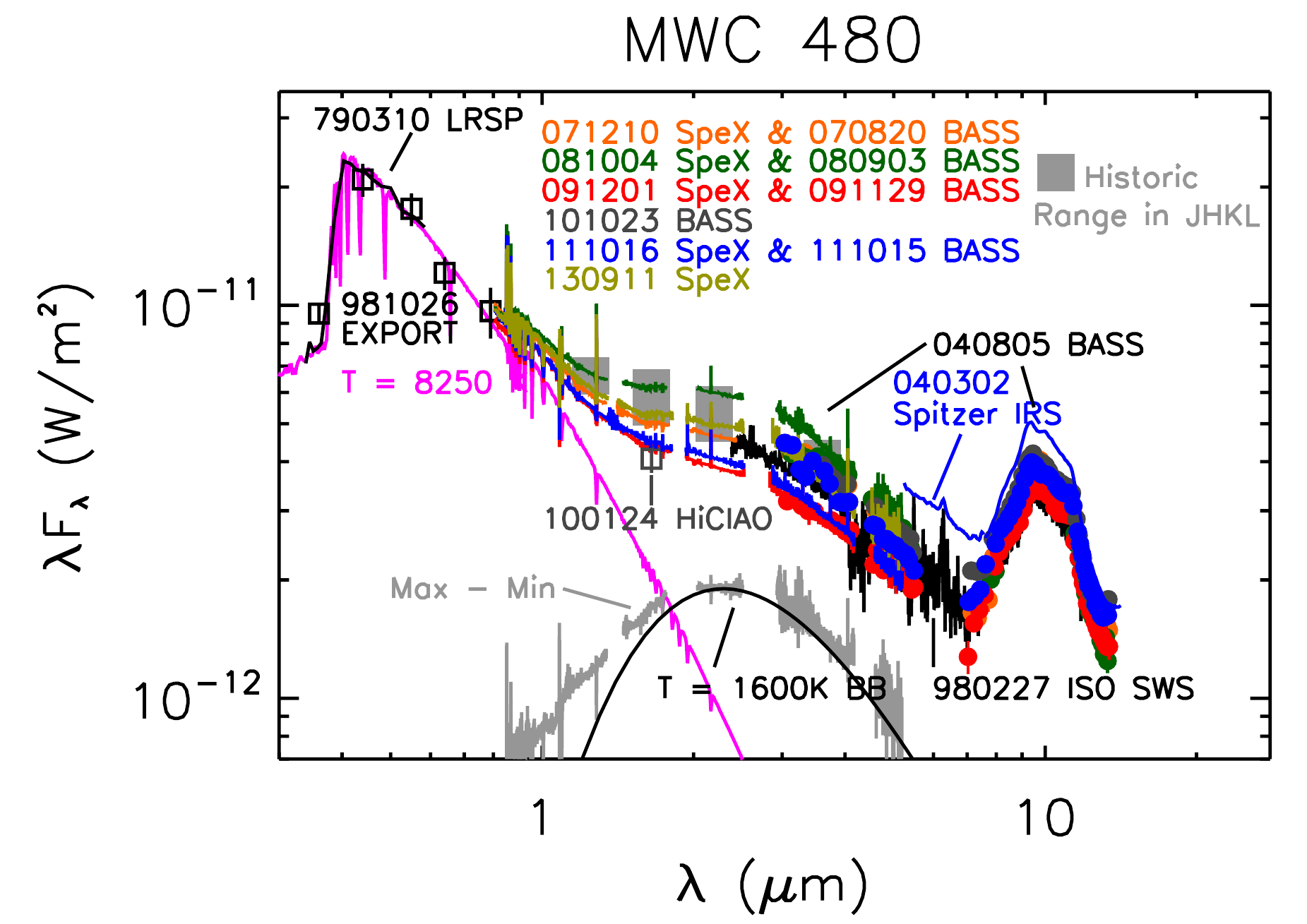}
\caption{The 0.34-13 $\mu$m spectral energy distribution of MWC 480. Multi-epoch data from BASS and SpeX, along with the 2010 HiCIAO photometry are shown. Also shown are UBVRI photometry from the EXPORT consortium \citep{Oudmaijer01}, data from the Low Resolution Spectrophotometer (LRSP) of the Pine Bluff Observatory \citep{Sitko08}, spectra from the Infrared Space Observatory (ISO) and the \textit{Spitzer Space Telescope}, and a T = 8250 K model atmosphere from \citet{kurucz79}. }
\end{figure}

\section{Mass Accretion Rates}
\subsection{Derivation of line Luminosities and mass accretion rates by Epoch}
The mass accretion rates were derived using the same procedures as those described by \citet{Sitko12}. The absolute flux-calibrated spectra were modeled as a combination of a stellar photosphere matched to the spectral type of MWC 480, plus a thermal component due to the hottest dust component in the system. For the spectral type, we adopted A3V \citep{Grady10}, and used SAO 206463, a pre-main sequence A0V star \citep{Houk82} star exhibiting neither significant gas accretion nor thermal dust emission for which we had comparable SpeX data. To this, we added a modified blackbody to approximate the thermal emission of the innermost dust. 
\\\\
Because the models were not able to produce a pseudocontinuum that matched the data in
every spectral order at every line to be extracted, the model was adjusted locally using a vertical
scaling until the $\chi^2$ of the difference in the continuum nearby (but outside) the line was minimized.
For the majority of the lines on all nights, these corrections to the scaling were less than 2$\%$ of the
initial model continuum level.
\\\\

\begin{figure}[H]
\figurenum{2}
\plotone{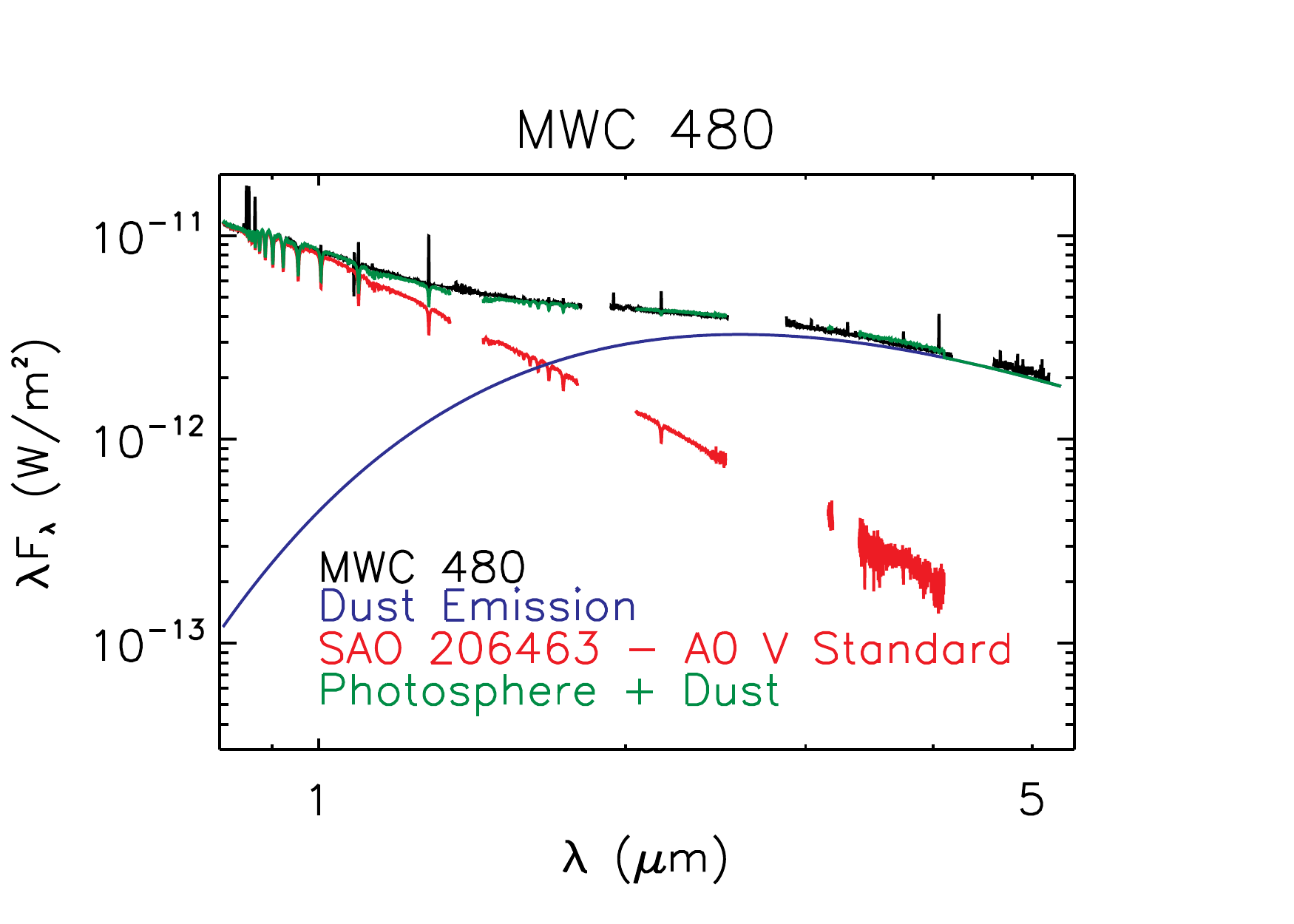}
\caption{Spectral ``model'' of MWC 480 used for extracting the line strengths of Pa $\beta$ and Br $\gamma$. Here the continuum flux (green curve) is modeled as a combination of a stellar photosphere (rad curve) and thermal emission by dust (blue curve) in order to best fit the actual SpeX data (black curve). For MWC 480, we used SpeX data on SAO 206463, a young A0V star with no known dust emission, and a modified blackbody, to roughly simulate a combination of a potential range in temperatures and wavelength-dependent dust emissivity. These were subtracted from the data on MWC 480 itself in order to extract the remaining gas emission. The ``model'' is not meant to by a physical one, but just a convenience to isolate the excess line emission from which mass accretion rates can be derived.}
\end{figure}

\begin{figure}[H]
\figurenum{3}
\plottwo{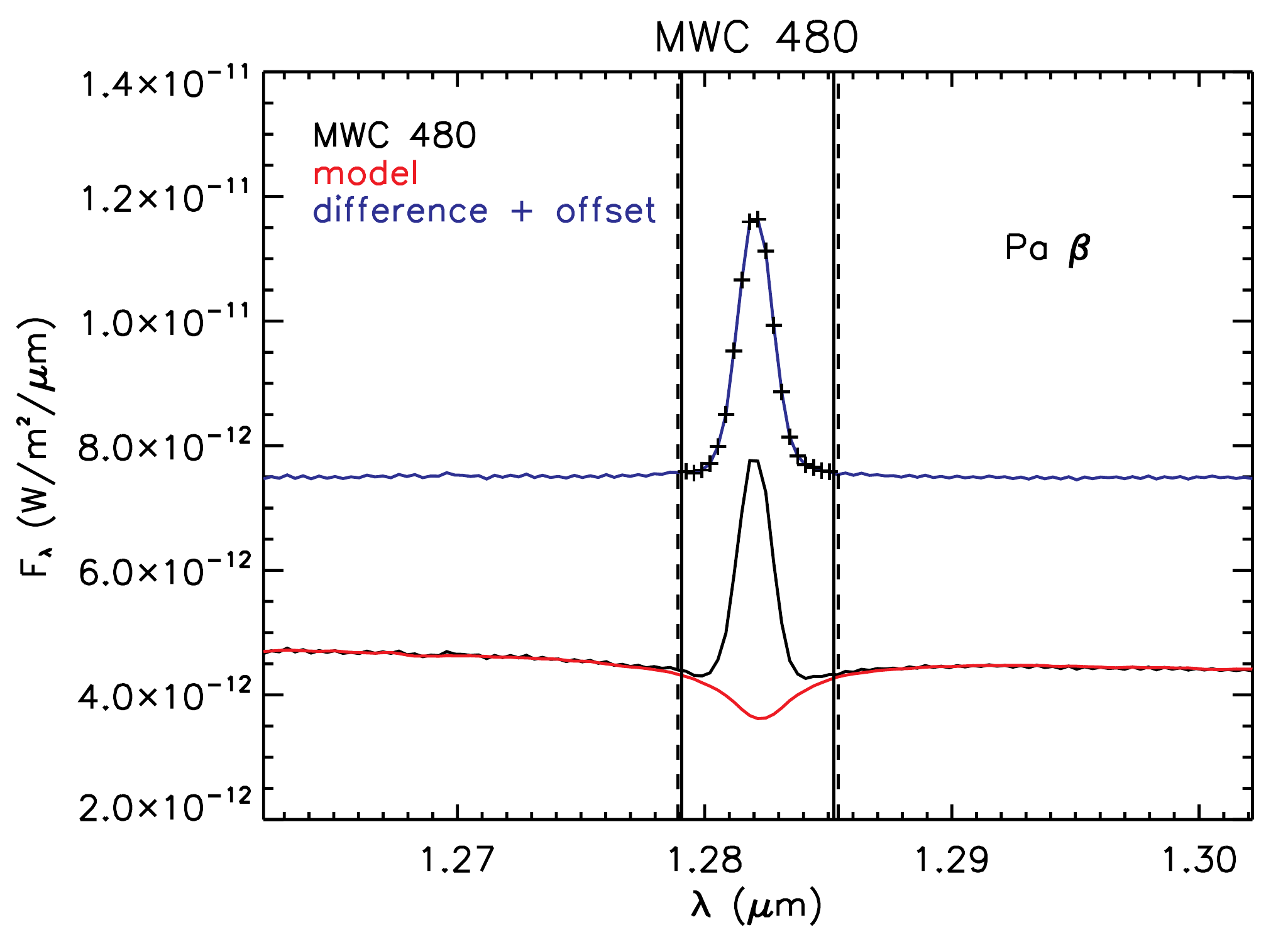}{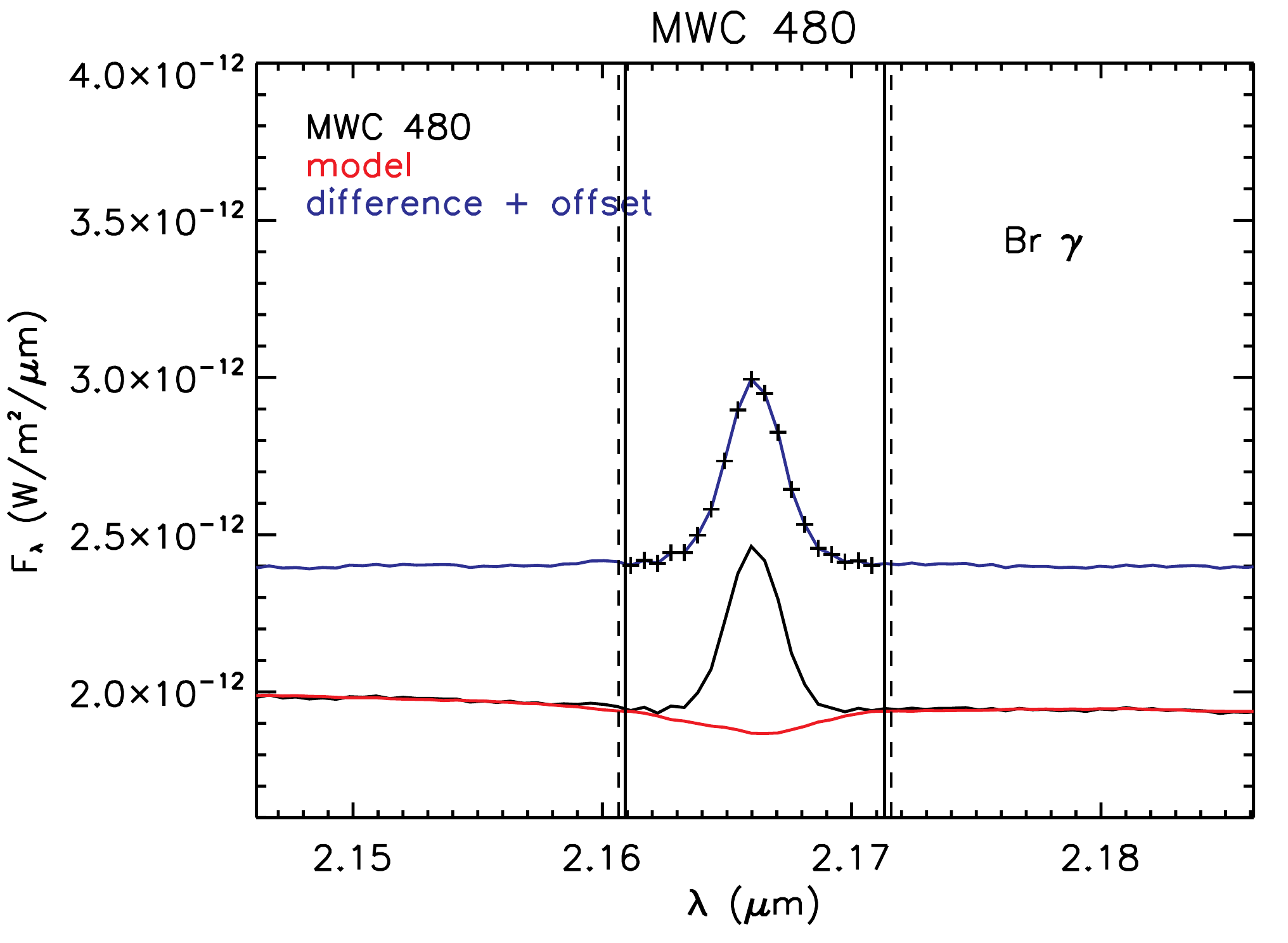}
\caption{Left: Extraction of the line emission in Pa $\beta$. The SpeX data is in black, along with the dust+photosphere in red (to improve the clarity in these line-extraction figures) , scaled to coincide outside the line. The upper blue curve is the difference, arbitrarily move upward to fit in the same figure (the wings are near zero). Right: The same for Br $\gamma$.}
\end{figure}

\vfill\eject

\begin{deluxetable*}{ccccc}
\tablecolumns{5}
\tablewidth{0pc}
\tabletypesize{\scriptsize}

\tablecaption{Br $\gamma$ and Pa $\beta$ Luminosities\tablenotemark{a} and Mass Accretion Rates\tablenotemark{b} in MWC 480}
\tablehead{
\colhead{Date (UT)} & \colhead{F$(Br \gamma)$} & \colhead{F$(Pa \beta)$} & \colhead{\.{M}$_{acc}(Br\gamma)$} & \colhead{\.{M}$_{acc}(Pa \beta)$} \\ (yymmdd) & (10$^{-15}$W m$^{-2}$) & (10$^{-15}$W m$^{-2}$) & (10$^{-8}$M$_{\astrosun}$y$^{-1}$) & (10$^{-8}$M$_{\astrosun}$y$^{-1}$)}
\startdata
071210 & 1.42 $\pm$ 0.07 & 7.22 $\pm$ 0.36 & 1.84 $\pm$ 0.21 & 2.04$\pm$ 0.24 \\
081004 & 1.79 $\pm$ 0.10 & 7.67 $\pm$ 0.46 & 2.48 $\pm$ 0.34 & 2.21 $\pm$ 0.31 \\
091201 & 1.47$\pm$ 0.08 & 6.79 $\pm$ 0.38 & 1.92 $\pm$ 0.25 & 1.89 $\pm$ 0.25 \\
111016 & 1.77 $\pm$ 0.09 & 7.83 $\pm$ 0.40 & 2.45 $\pm$ 0.29 & 2.77 $\pm$ 0.27 \\
130911 & 1.14 $\pm$ 0.06 & 7.22 $\pm$ 0.41 & 1.39 $\pm$ 0.19 & 2.044 $\pm$ 0.27 \\
\enddata

\tablenotetext{a}{ ~The line luminosities are based on a distance of 142 pc. The conversion of line luminosities to accretion luminosities uses the relationships derived by \citet{Fairlamb17} for Pa $\beta$ and Br $\gamma$. These are derived for lower accretion rates and may suffer from systematic uncertainties. The quoted uncertainties do not include that of the original calibration. An added systematic uncertainty is present due to the 5\% uncertainty in the distance, resulting in uncertainties in the line luminosity, stellar radius, and stellar mass, combined to produce an uncertainty in the accretion rate of $\sim$13\%. These do not include the uncertainties of these two lines from \citet{Fairlamb17}, where, for example, the zero-point uncertainties are 0.16 dex and 0.23 dex for Pa $\beta$ and Br $\gamma$, respectively. }
\tablenotetext{b}{ ~Using \.{M} $\approx$ L$_{acc}$R$_{\ast}$/GM$_{\ast}$ with R$_{\ast}$ = 1.96 R$_{\astrosun}$ and M$_{\ast}$ = 2.15 M$_{\astrosun}$.}

\end{deluxetable*}

\vfill\eject

The line strengths of the Pa$\beta$ and Br$\gamma$ lines were then extracted by subtraction of the adjusted model from the flux-calibrated data, and the net line flux calculated by integrating over the flux-calibrated continuum-subtracted line profile. These were converted to line luminosities using a distance of 142 pc. The line luminosities were transformed to mass accretion luminosities using the calibrations of \citet{Fairlamb17} for Pa$\beta$ and Br$\gamma$, respectively. These were then converted to mass accretion rates using \.{M}= {L}$_{acc}$R$_{\ast}$/GM$_{\ast}$, with adopted values of R$_{\ast}$ = 1.93 R$_{\astrosun}$, and the stellar mass M$_{\ast}$ =2.15 M$_{\astrosun}$. While there are many different calibrations of the mass accretion rates based on these line, we selected that of \citet{Fairlamb17}, as it was derived specifically for Herbig AeBe stars, and has among the smallest calibration uncertainties in the literature. The mass accretion rates for MWC 480 were time-dependent, being (1.43 - 2.61)$\times$10$^{-8}$M$_{\astrosun}$y$^{-1}$ and (1.81 - 2.41)$\times$10$^{-8}$M$_{\astrosun}$y$^{-1}$ derived from Br$\gamma$ and Pa$\beta$ lines, respectively.

\subsection{Line versus Continuum Emission}

One aspect of the continuum and line variability that has generally received little attention is whether these two different sources of flux are in any way related.  \citet{Sitko12} illustrated the lime emission strengths of Pa $\beta$, Br $\gamma)$, and the O I line at 0.8446 $\mu$m versus the flux in the K band in SAO 206462. At the start of the increase in K band flux in 2009, the strengths of all three lines were observed to increase, and continued to do so after the continuum flux receded in strength. This might indicate that whatever caused the warm dust emission to increase triggered some sort of response in the gas accretion. However, in the case of HD 163296, \citet{Sitko08} found little change in the Paschen lines between the two epochs with the largest change in NIR continuum yet seen in that object. The Ca II lines actually became stronger as the continuum decreased.\\

For MWC 480, the SpeX data were obtained at intervals of 1 to 2 years apart, not close enough in time to make a meaningful light curve. In this case we simply examined whether the gas and dust emission were in any way related to one another. The result is shown in Figure 12 where the Pa $\beta$ and K band fluxes are shown.  It is clear that a more concerted effort to monitor MWC 480 will be required to investigate the possible connection between line and continuum emission.

\begin{figure}[H]
\figurenum{4}
\centering
\includegraphics[scale=0.7]{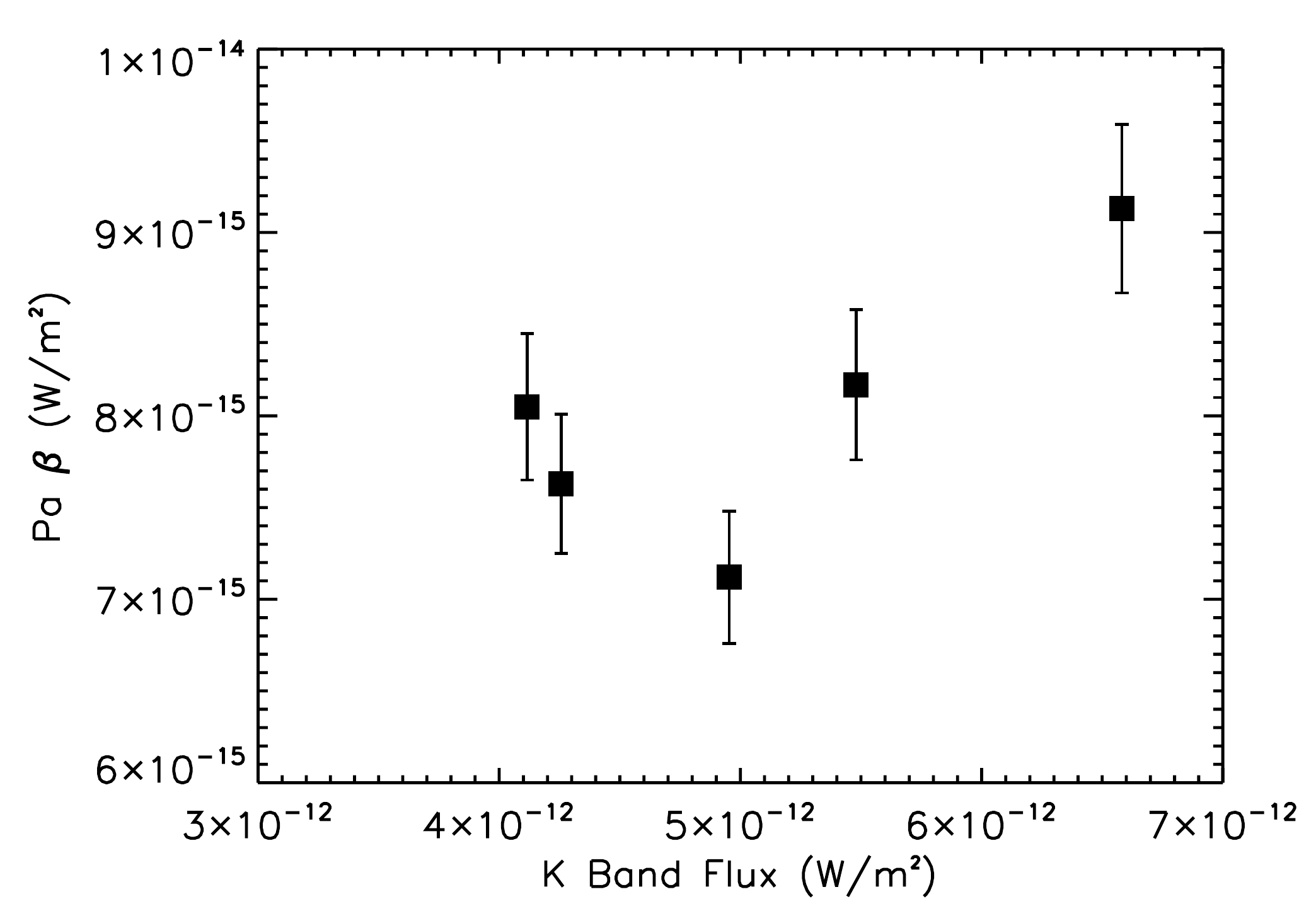}
\caption{The flux in Pa $\beta$ versus the flux in the K band. While, unlike SAO 206462 \citep{Sitko12} no strong correlation is apparent, and a larger sample of observations obtained closer in time are needed to investigate any possible relation.}
\end{figure}

\section{Radiative Transfer Modeling}
In order to model the SED of MWC 480, we used the three-dimensional Monte Carlo Radiative Transfer (MCRT) code HOCHUNK3D of \citet{Whitney13}. HOCHUNK3D allows for two independent, constituent coplanar disks, each having its own set of input parameters which dictate the structure and composition of that disk. These include different vertical scale heights to simulate grain growth and settling \citep{Dullemond04a, Dullemond04b} as well as different radial extents. We assigned one disk to contain ``large" grains with a small scale height (``settled disk'') and the other  disk to contain smaller grains with a larger scale height that overlaps the first disk. Our model includes a third component that was originally developed for an envelope of material infalling onto the disk, but that we employed to create an inner disk ``fan'', as an approximation of a disk wind.\\  

The actual radial dimensions of the disk surrounding MWC 480 can be characterized by the size of the disk observed in scattered light from the dust,  the size determined through the thermal emission of the dust, and through the emission by the molecular gas. \citet{Pietu07} traced CO emission to distances of 700-800 au. However \citet{Pietu06}, using a 0.73 x 0.53 arcsec beam,  traced the millimeter continuum emission by the dust to only 185-190 au, indicating that either the dust does not extend as far as the gas (e.g. \citet{birnstiel14}), or that it was simply not detectable. \citet{huang17}, using the Atacama Large Millimeter/submillinmeter Array (ALMA), found the outer radius of the dust disk to be 200 au, in agreement with the value of  \citet{Pietu06}.  \citet{Kusakabe12}  traced the scattered light out to only 137 au, using the Hight Contrast Instrument for Subaru Next Generation Adaptive Optics (HiCIAO) instrument on the Subaru telescope (the pixel scale was 9.53 mas/pixel, and the PSF FWHM was measured to be 0.07 arcsec.). Given the faintness of the disk (it has largely eluded ``strong'' detection in the past), it is possible that the instrument was not sufficiently sensitive to detect the disk further out. It might also be the case that the disk is becoming self-shadowed at larger distances.  Since our study deals primarily with the thermal emission between 1 $\mu$m and 100 $\mu$m (the latter has characteristic temperatures $\sim$ 30-40 K and for an A3V star, distances $\sim$60 au), and of the scattered light, the 200 au radial size for the disk measured with ALMA was adopted.\\

The inner radius of the disk is somewhat difficult to establish. Using the Keck Interferometer Nuller operating in the N band (8.0-13.0 $\mu$m), combined with SED data from 0.5-13.0 $\mu$m, \citet{Millan-Gabet16}  modeled the interior region of MWC 480 two ways - one model with a single inner rim and a second model with two inner rims. In the model consisting of a single inner rim, the inner radius was found to be 0.10 au, and had a characteristic temperature of 2500 K. As this greatly exceeds typical sublimation temperatures for silicates ($\sim$1500 K), such a system would require either super-refractory grains or hot gas at those distances (see, for example, \citet{Benisty10} for a discussion for the case of HD 163296).  In the two-rim model, they derived distances of 0.44 au and 2.3 au (both with inner rim temperatures below 1500 K). \citet{Lazareff17} observed MWC 480 using the PIONIER instrument on the VLT in the H band (1.55-1.80 $\mu$m). Aided by SED information, they found that an ellipsoid with a Half Width Half Maximum (HWHM) of 1.9 au provided the best fit. While a ring model was determined for many objects in their study, none was found for MWC 480. Due to the uncertainties presented by the inteferometrically-determined dimensions, we placed an emphasis on using the SED in 1-5 $\mu$m as a guide. Nevertheless, we recognize that 1-2 $\mu$m gas emission, which is not included in the HOCHUNK3D code, is likely to cause additional emission in this wavelength. \\ 

Here we investigate two different scenarios for the observed variability of the NIR fluxes derived from the SpeX data. In the first one, we use disk with an inner rim, the scale height of which changes in a manner as to produce the changing NIR fluxes. In the second case, we change the structure and density of an inner wind or jet. 

\subsection{Changing the puffed up inner rim}

We began this investigation by setting the ``low'' NIR photometric state as the starting point of the modeling. Fitting the low state required setting the inner rim height to 0.05 au, with an added 50\%  rim ``puff'', increasing it to 0.075 au, which reproduced the observed SED. A plot of the temperature and density in this model are shown in Figure 1, and the resulting SED in Figure 2. As is apparent in Figure 2, this model reproduces the NIR low flux state, as well as the remainder of the available data at longer wavelengths. \\

To fit the high state, we used the same model, but raised the value of the inner rim height just enough to reproduce the higher NIR flux levels in the SpeX data (increasing the rim ``puff" to 50\%). The result of this change is shown in Figure 3. As is evident in the resulting SED, the ``see-saw'' behavior is reproduced, with the model fluxes at longer wavelengths dropping accordingly. The $\sim$30 \% change in the NIR flux produces a more substantially visible change at longer wavelengths, since the total fluxes are originally more than an order of magnitude lower to begin with. Figure 4 shows a  side-by-side comparison of the inner disk region for the two models .\\

More importantly, while raising the height of the inner wall fits the NIR region well, it produced far-IR emission that was substantially lower than all observed flux levels. While there are no far-IR observations that are nearly simultaneous with any of the NIR data, the fact than \textit{none} of the data at longer wavelengths can be fit by this second model suggests that simply increasing the rim height is unlikely to be the source of the observed NIR variations. We checked the reasonableness of this result by determining the change in surface brightness in the outer disk produced by this change in inner rim scale height. We measured the azimuthally-averaged surface brightness in the H band, using a face-on geometry to minimize the effects of disk inclination and disk shape, to provide an indication of just how much change there was in the light actually reaching the outer disk. In these models, the surface brightness, shown  in Figure 5, dropped as $\sim r^{-2.0}$, but was roughly only one-half the brightness in the high-rim state as the low-rim state. The relative change is less than what the observer sees from the inner rim, and the disk ``sees'' the inner regions at higher inclination, and in such a thin disk, even a small change in inner rim height can create a substantial deficit of light from the star reaching the outer disk. One way such differences in light reaching the outer disk could be masked is the delay in any ``thermal pulse'' from propagating deep onto the disk, effectively smoothing thermal emission response. While it is possible that the time scale for such a ``thermal pulse'' to propagate into the disk mid-plane will be years (see the discussion in \citet{Sitko12}), it is evident that out to at least 37 $\mu$m many disk system show measurable variability, with ``see-saw'' response \citep{Espaillat11}, which seems to be absent in MWC 480. This, and the presence of a jet in imaging observations of MWC 480, suggest that an alternative model be explored.\\

\begin{figure}[H]
\figurenum{5}
\centering
\includegraphics[scale=0.5]{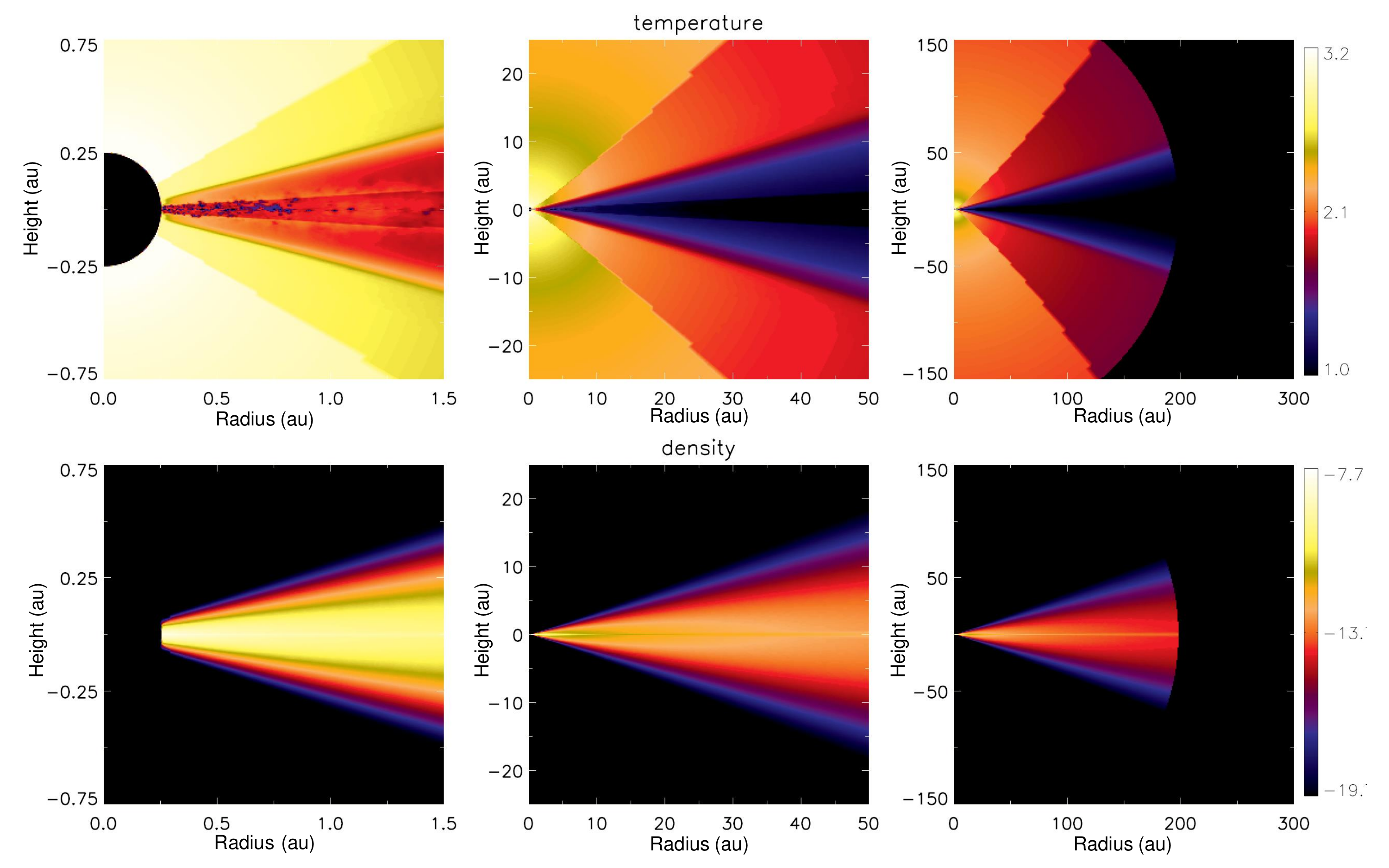}
\caption{The horizontal and vertical temperature (upper three panels) and density (lower three panels) distribution in the circumstellar material of the high NIR state, modeled with just a slightly puffed up inner rim. The horizontal and vertical scales are in au, with, from left to right, the the structure from 0-1.5 ua, 0-50 au, and 0-300 au. The temperature scale color bar is the log of the temperature in Kelvins, while the density scale is the log of the mass density in g cm$^{-3}$.} 
\end{figure}

\begin{figure}[H]
\figurenum{6}
\centering
\plottwo{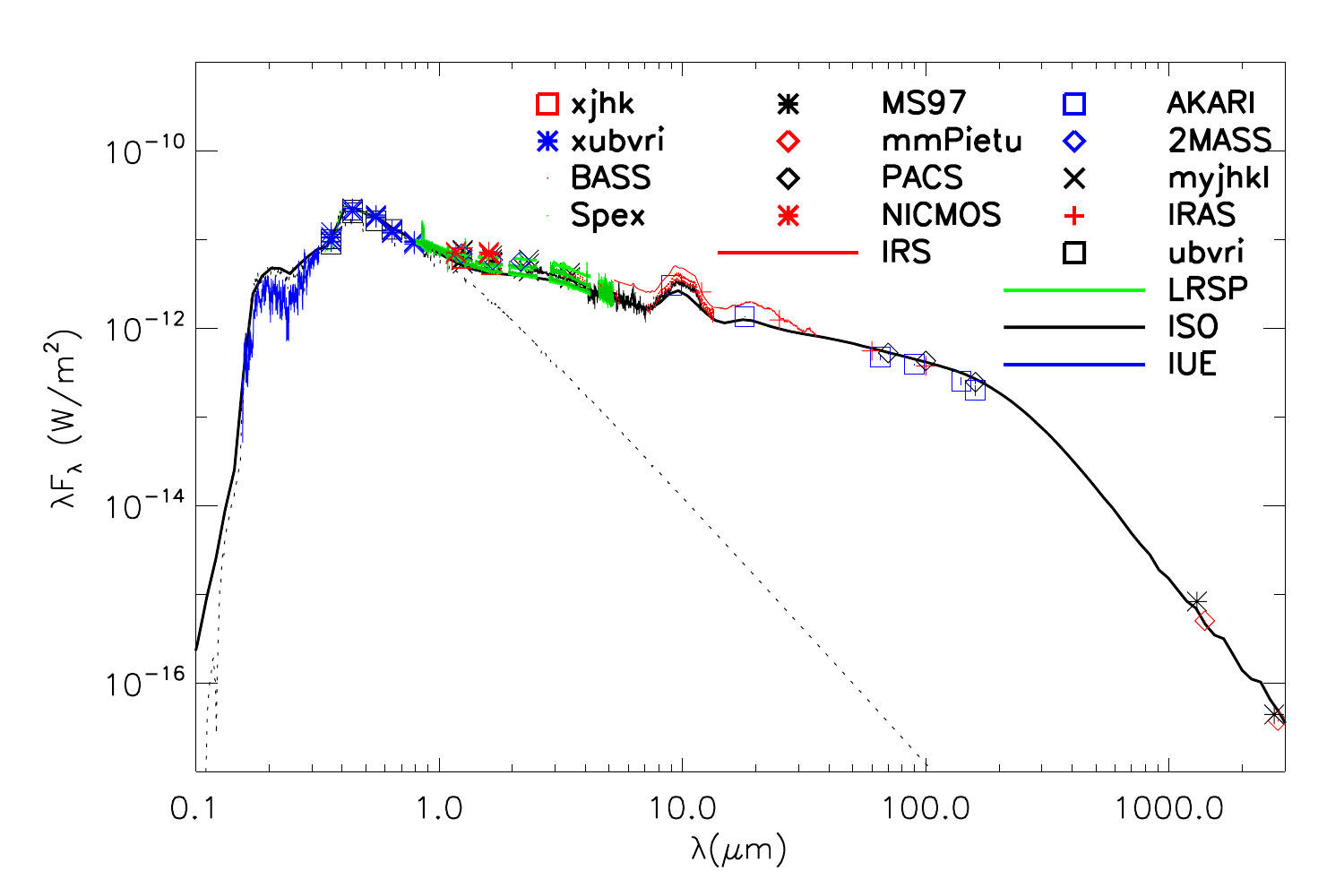}{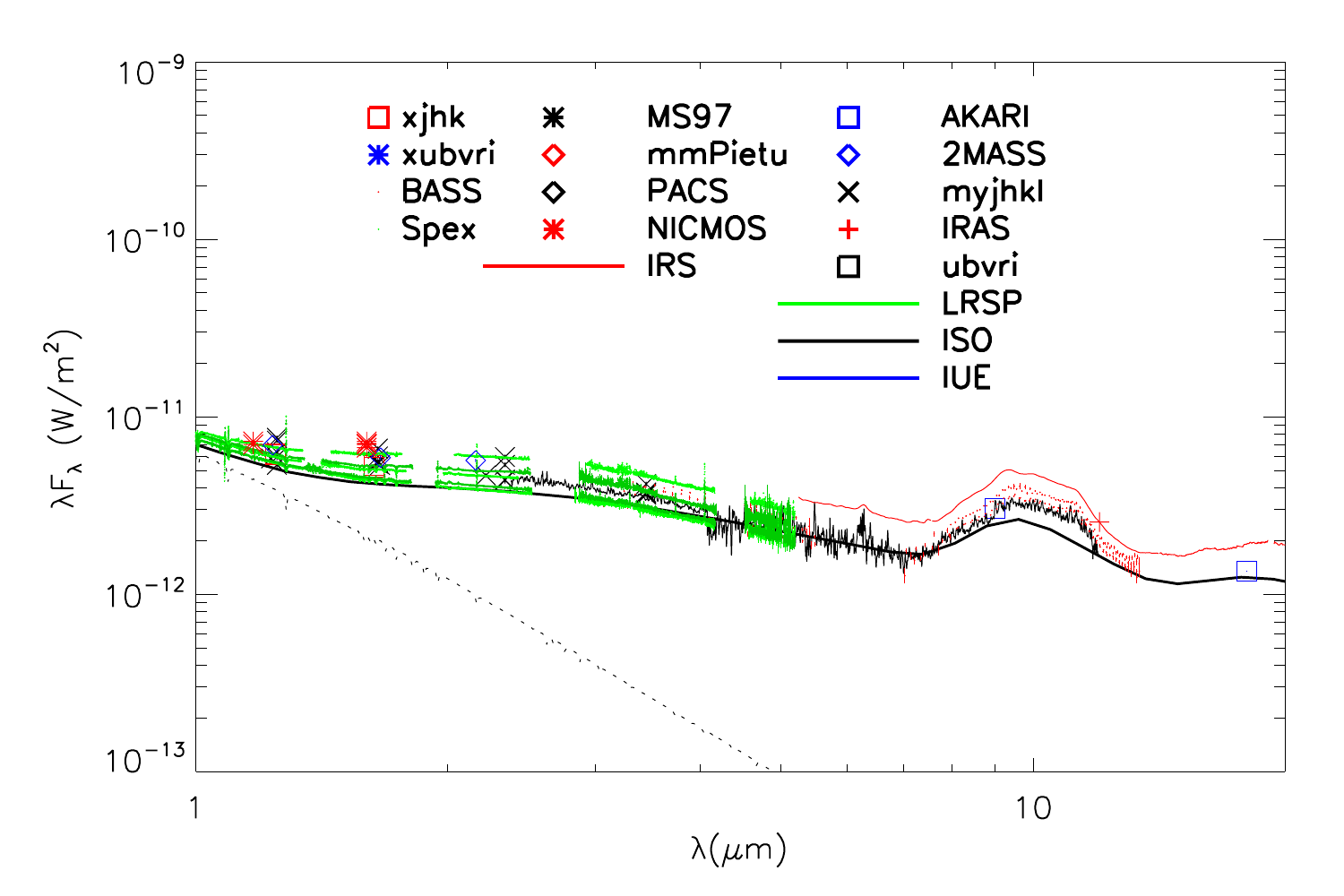}
\caption{Left: SED created to fit the low state NIR data and the data at longer wavelengths. In the region of the 5-12 $\mu$m region, the Infrared Space Observatory Short Wavelength Spectrometer (ISO SWS) data shown characterizes the low state, but the longer-wavelength spectrum was not included, due to its low quality. By contrast, the Spitzer IRS spectrum was apparently obtained during a higher NIR flux state than all of the other near- to mid-IR data sets. Right: A closeup of the near-IR data and the same model fit. Note that the dates of observation for the ISO SWS and \textit{Spitzer} IRS were 27 February 1998 and 2 March 2004 UT, respectively, and are outside the range of epochs of the SpeX and BASS observations. }
\end{figure}

\begin{figure}[H]
\figurenum{7}
\centering
\plottwo{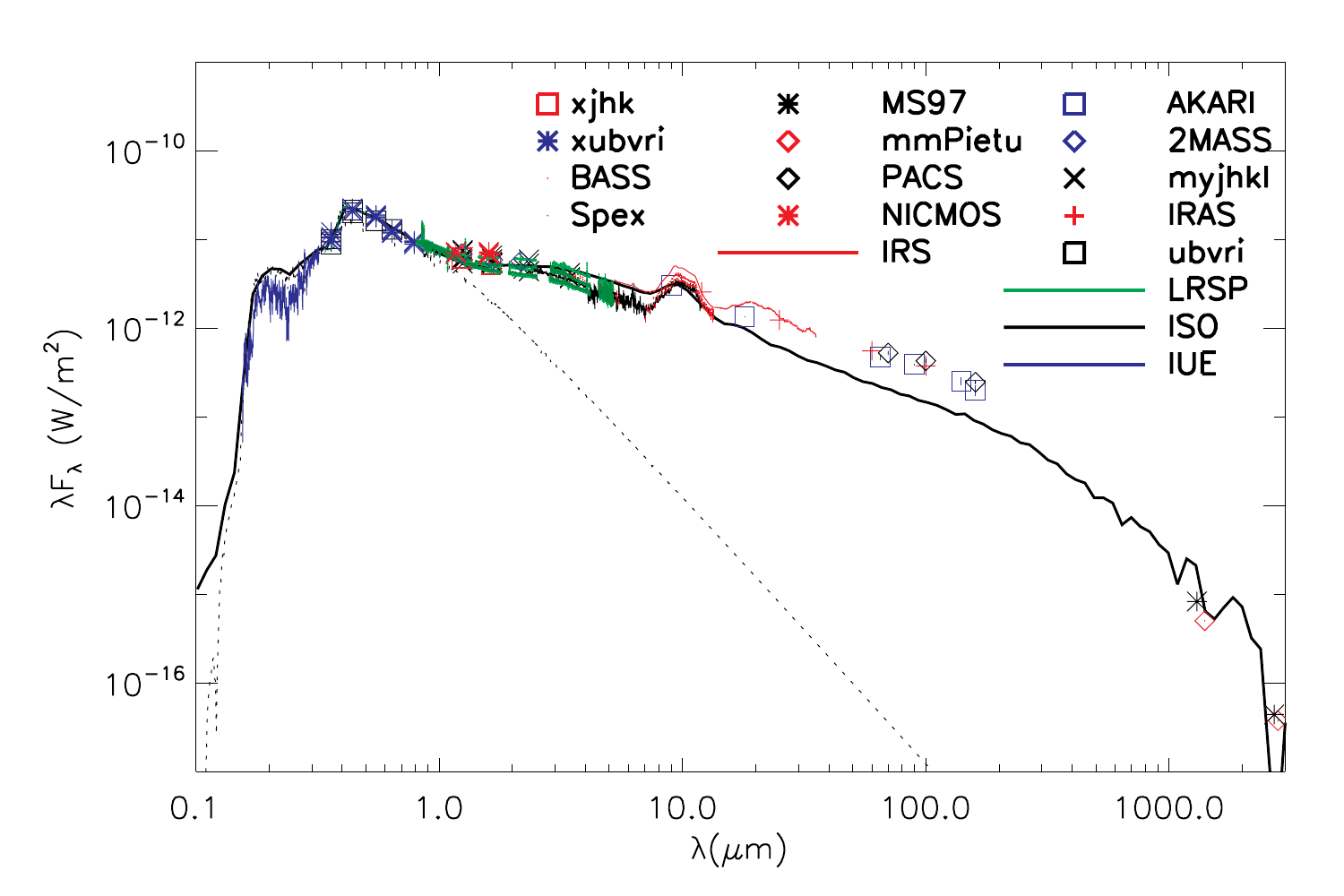}{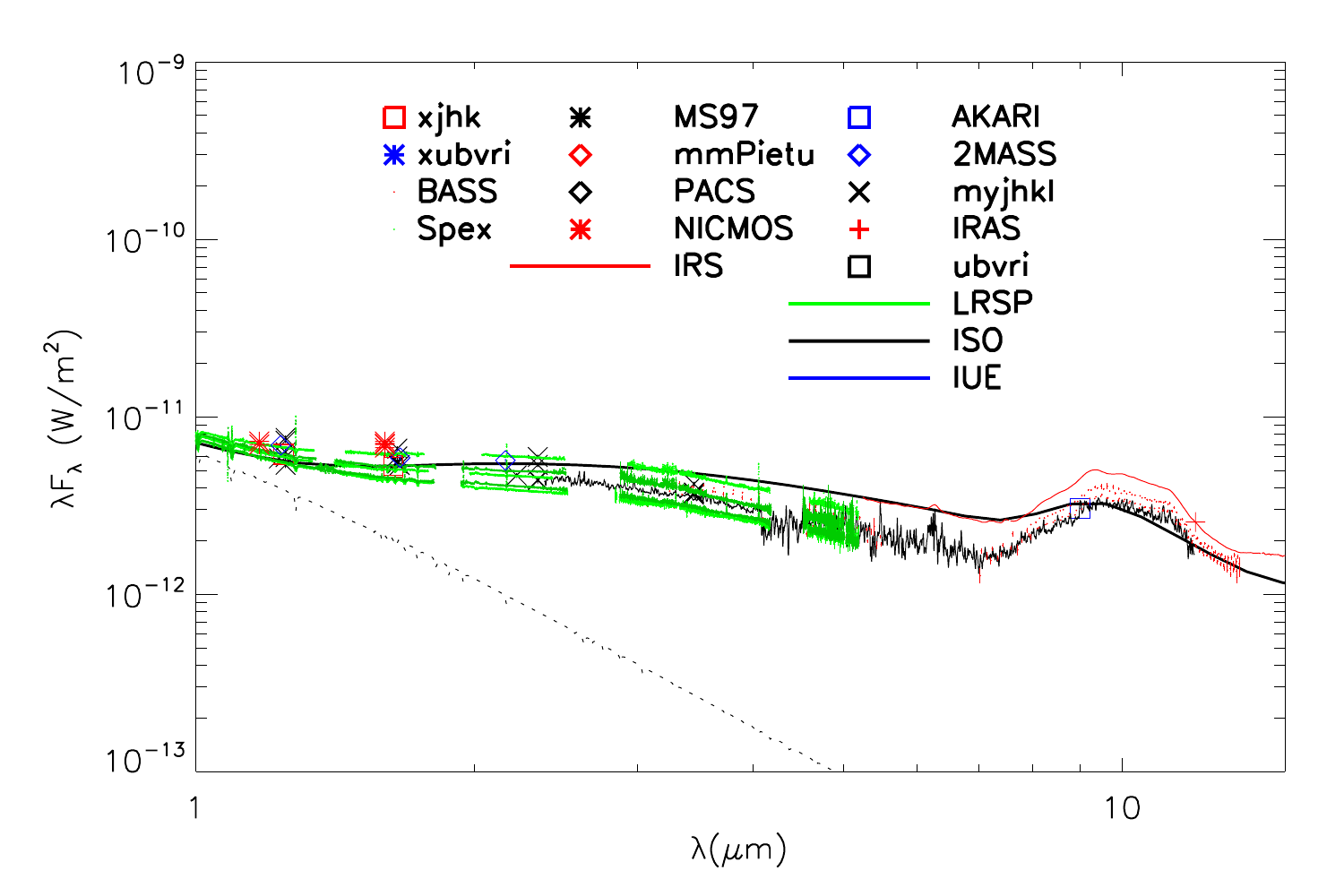}
\caption{SED and model of the high NIR flux state, produced by increasing the scale height of the puffed up inner disk rim. With a modest change in the height of the puffed up rim,  the model produces far-IR emission substantially lower that all observed flux levels. Right: Closeup of the same data and model. It successfully reproduces the "see-saw" behavior observed in some disk systems, but such behavior is inconsistent with the available data at longer wavelengths.}
\end{figure}

\begin{figure}[H]
\figurenum{8}
\centering
\plottwo{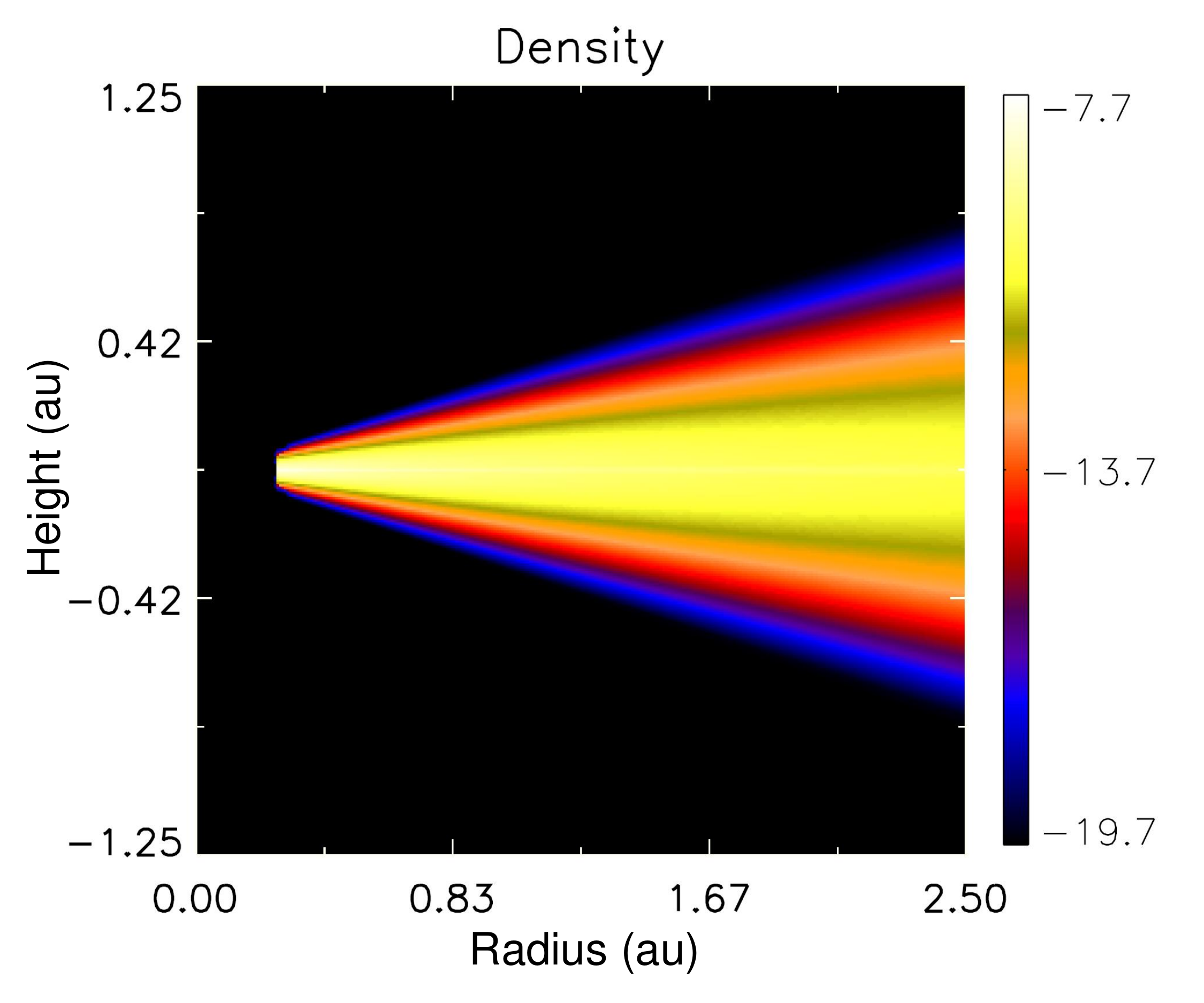}{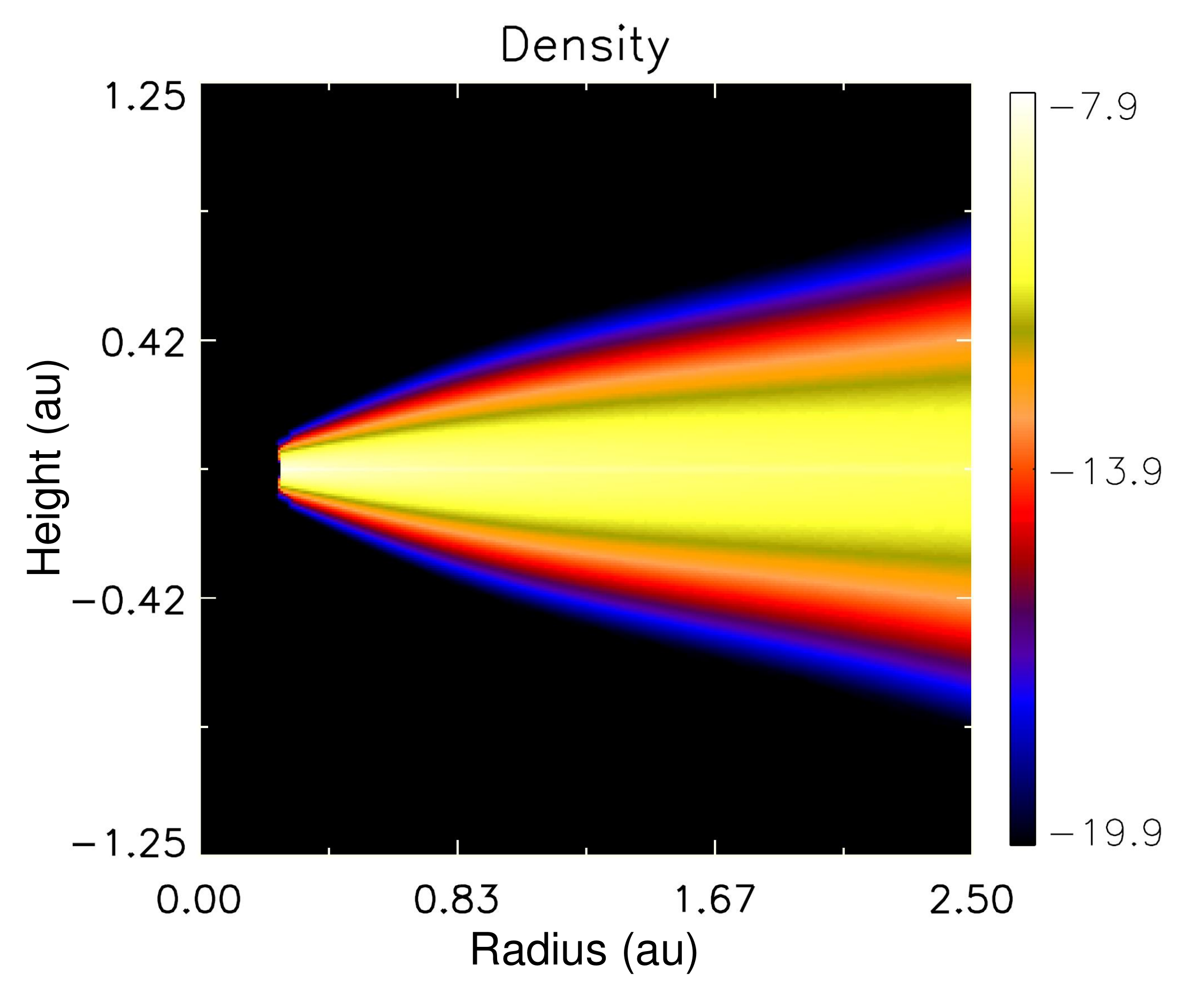}
\caption{A comparison of the mass density inner disk region for the NIR low state (left) and NIR high state (right).}
\end{figure}

\begin{figure}[H]
\figurenum{9}
\centering
\includegraphics[scale=0.7]{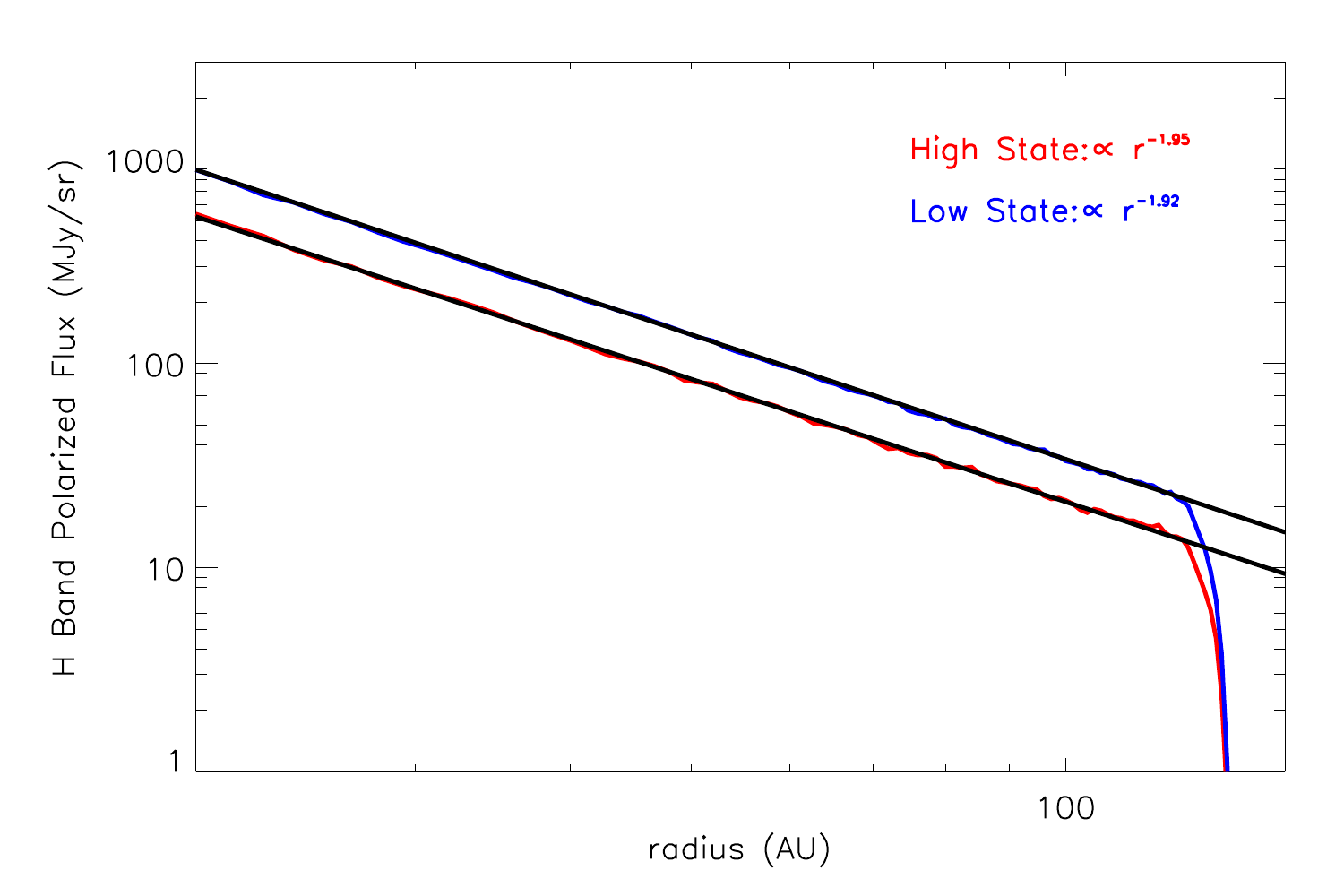}
\caption{The surface brightness versus radial distance for the variable disk rim height models. Here, rather than deriving the surface brightness of the curved upper disk surface seen by the observer at the inclination of the disk of MWC 480, which will suffer significant distortions, we show the effective scattering of the disk, as seen face, in order to remove the distortion produced by the observed geometry, and allowing azimuthal averaging of the signal. In this manner. Also shown in the upper right corner of the figure is the radial dependence of the scattering. The upper curve represents the ``low'' inner rim state, which allows more stellar light to reach the outer disk, while the lower curve represents a higher inner rim ``puff'' that reduces the amount of light reaching the outer disk. A ``low'' rim state might explain the greater ease of detection compared to previous attempts, but such a change should also change the mid-IR portion of the SED, for which there is no evidence yet.}
\end{figure}

In this pair of models, any increase in the size of the inner rim, which is generally envisioned to be optically thick (in order to reproduce the ``see-saw'' behavior observed in some other disks) significantly reduces any direct illumination by the star, plunging the outer disk into darkness, and dropping its emission below all observed levels. Even with non-simultaneous mid- and far-IR archival data sets we would expect to see scatter in the data comparable to the difference between the observed and model flux levels shown in Figure 3. This suggests that this scenario is not likely to explain most of the near-IR variability seen in MWC 480.\\

The emission of the inner wall requires a combination of both optical depth and physical scale. An optically thick inner wall can produce the observed NIR emission with a minimum solid angle, as ``seen'' by the illuminating star. Lower optical depth requires a larger solid angle structure. Because MWC 480 is known to drive  bipolar jets, we investigated an alternative geometry where the NIR is dominated by a more optically thin structure. \\

\subsection{Changing the inner disk wind structure}

For the jet/wind models, we began again with a goal of first reproducing the low flux state in the NIR, and the remaining observations at longer wavelengths. The temperature and density structures of this model is shown in Figure 6, while the resulting model SED is in Figure 7. Changing only parameters relevant to the inner few au, that is by increasing the ``ffducial'' density of the wind component at 1 au from 2.5$\times$10$^{-16}$ g/cm$^{3}$ to 5$\times$10$^{-16}$ g/cm$^{3}$ (this is a basic model parameter in the code, and is independent of the \textit{actual} inner radius of the model) , increasing the exponent of power law mass density dropoff with distance $r$ from $\rho \sim r^{-1.2}$ to $\rho \sim r^{-1.8}$, and pushing the actual inner envelope gap radius from 1 au to 2.5 au, we reproduced the high NIR flux state, as shown in Figure 8. A closeup of the density in the inner regions of the two models is shown in Figure 9.  As is apparent in the model SEDs of the wind/jet models, the agreement between both models and the flux levels at longer wavelengths is better than that of the high flux NIR state of the changing inner rim models. Thus, this scenario would seem to be consistent with the ensemble of data sets that are available.

\begin{figure}[H]
\figurenum{10}
\centering
\includegraphics[scale=0.5]{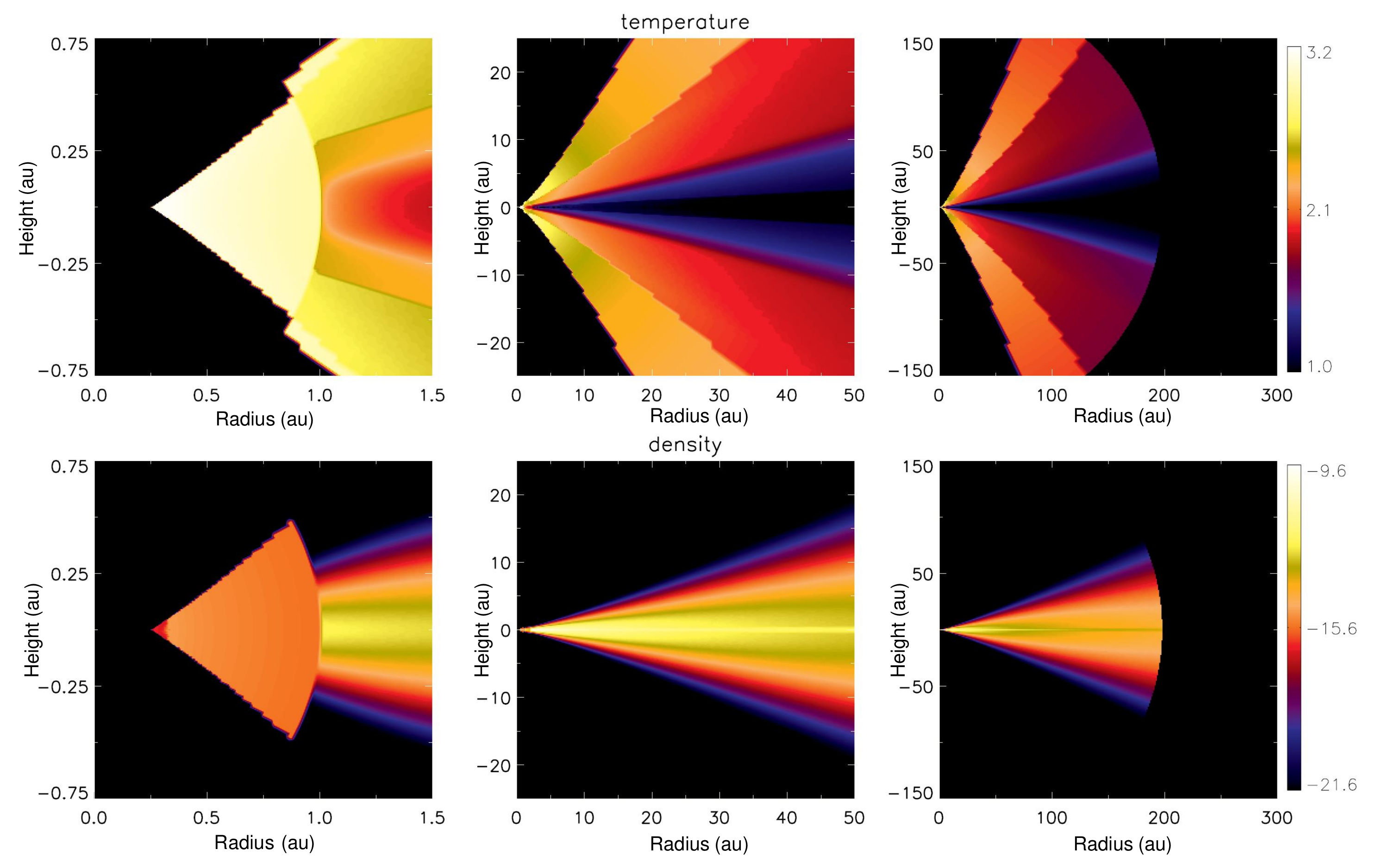}
\caption{The above figure shows the temperature and density distribution of the low NIR state, modeled with just a slightly puffed up inner rim but with material above and below the disk, representing the denser grains of a disk wind.}
\end{figure}

\begin{figure}[H]
\figurenum{11}
\centering
\plottwo{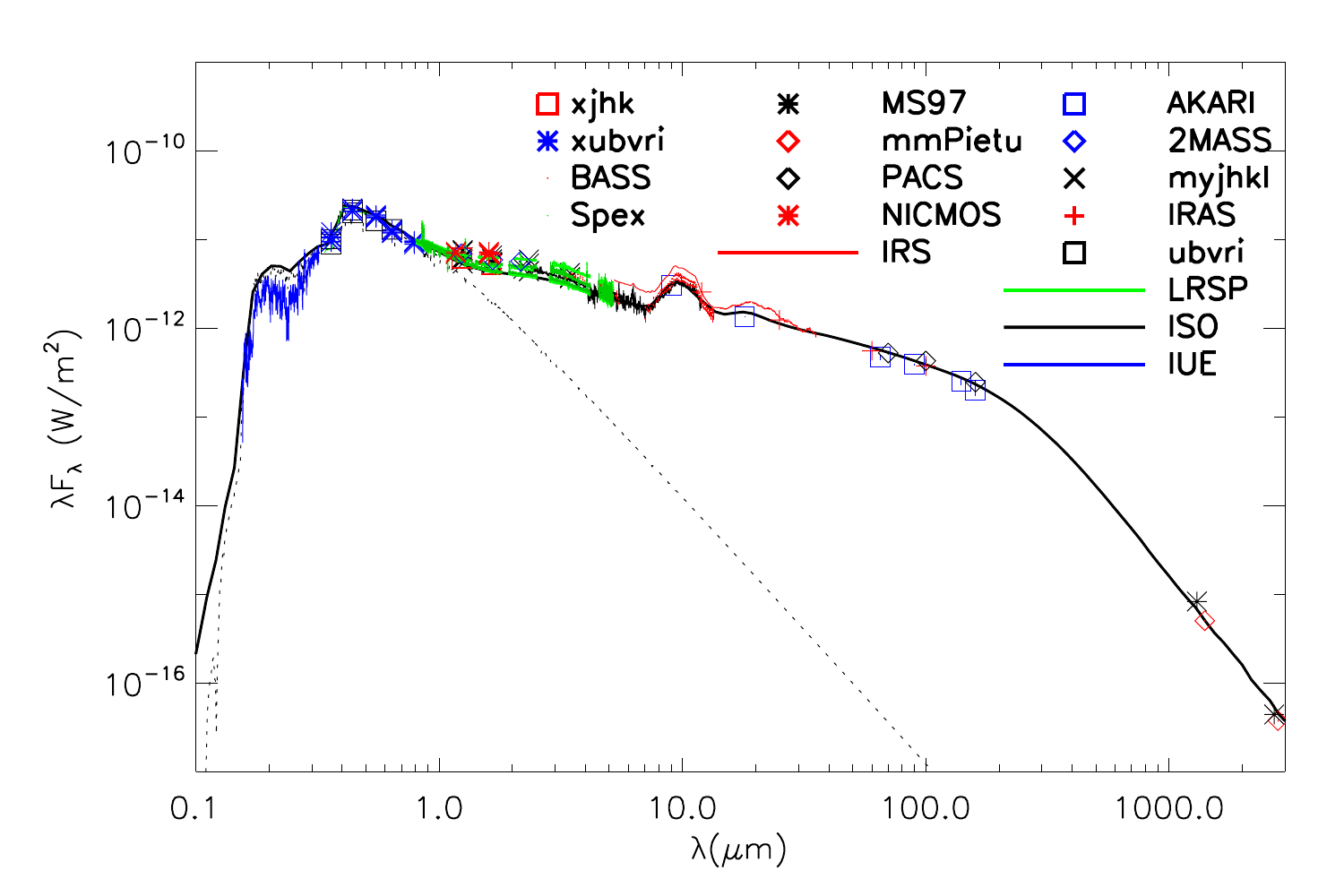}{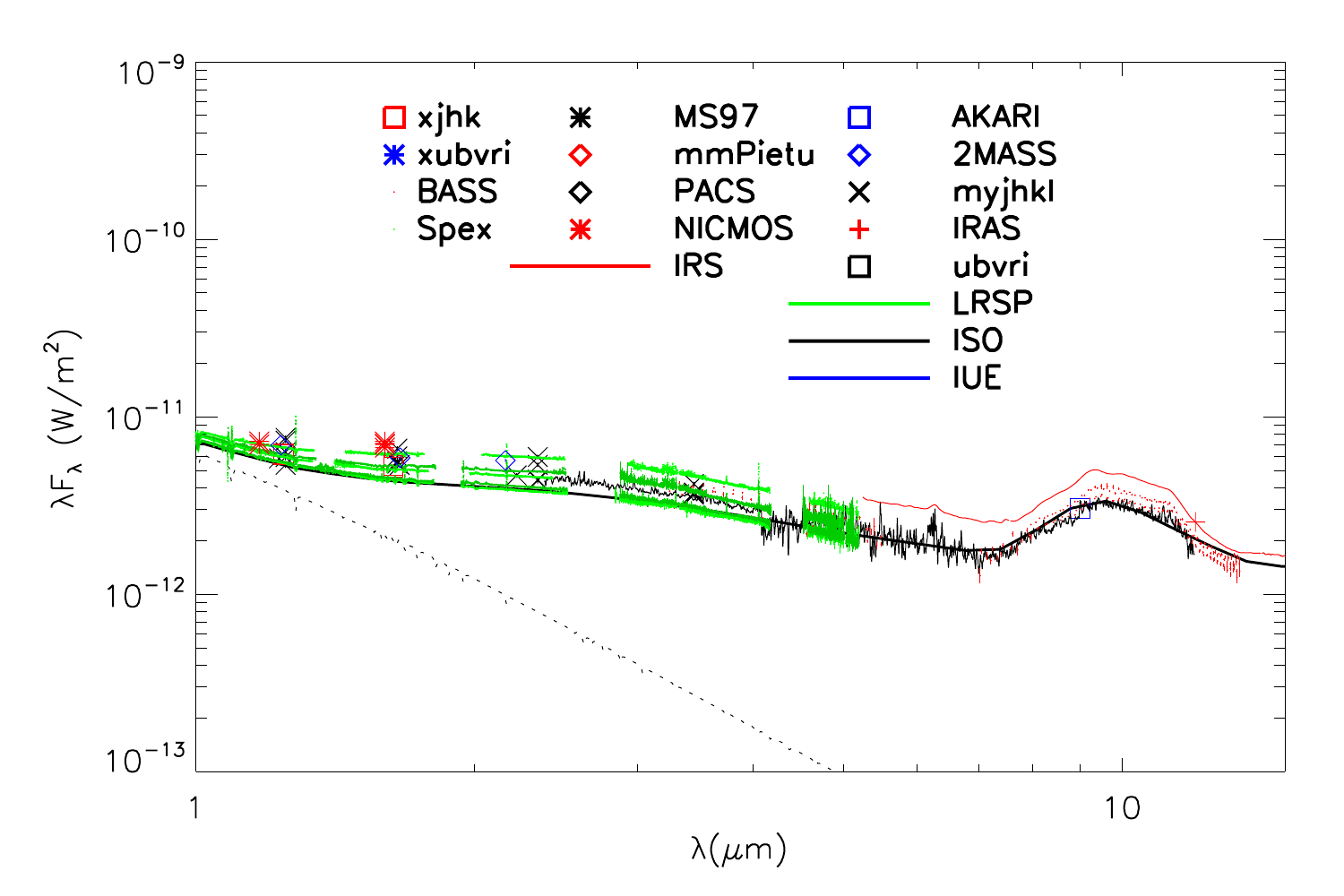} 
\caption{Left: SED of the low flux state, this time fit using a model with a ``disk wind'' geometry. This was created to fit the far-IR data as well. Right: A closeup pf the same data ad model.}
\end{figure}

\begin{figure}[H]
\figurenum{12}
\centering
\plottwo{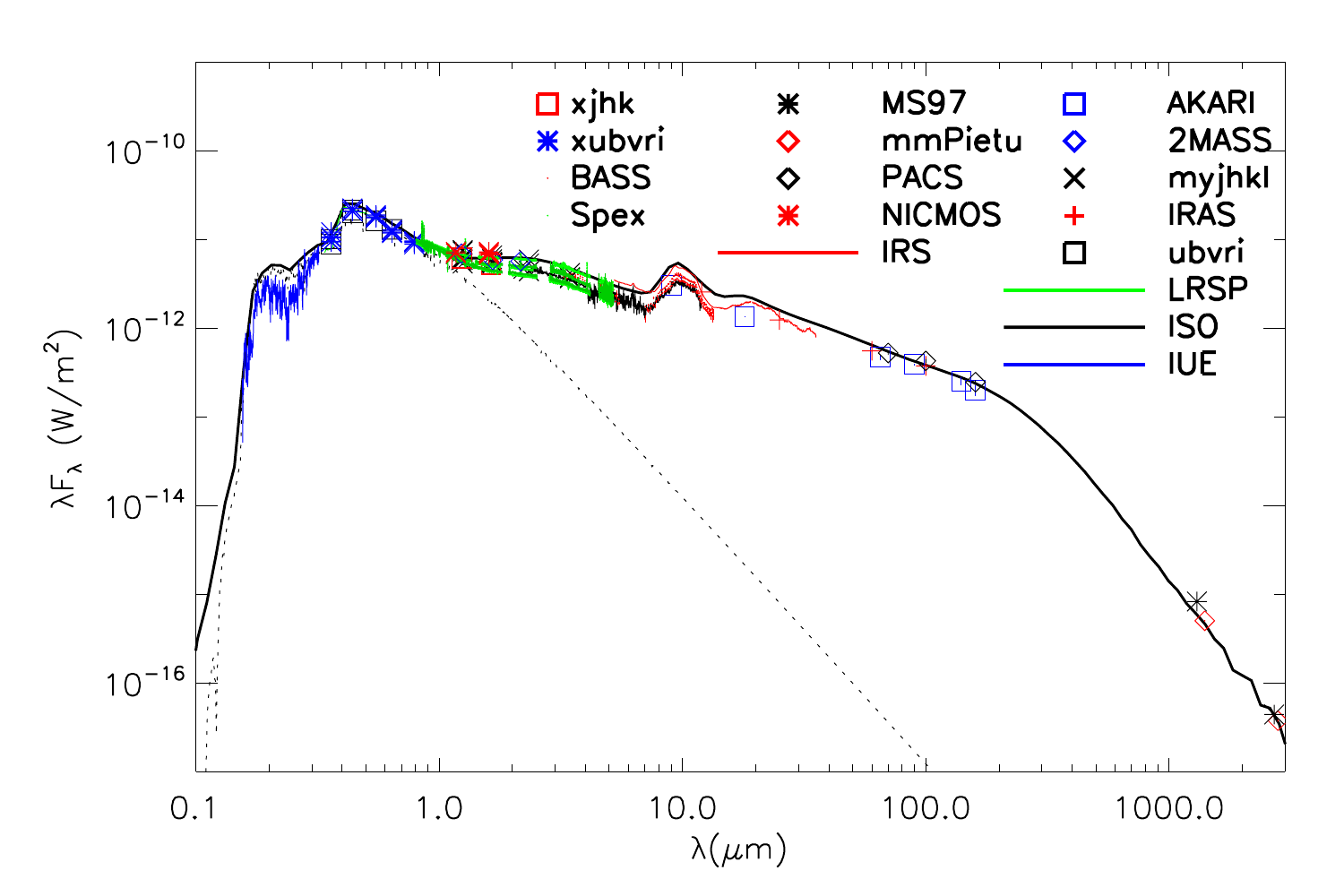}{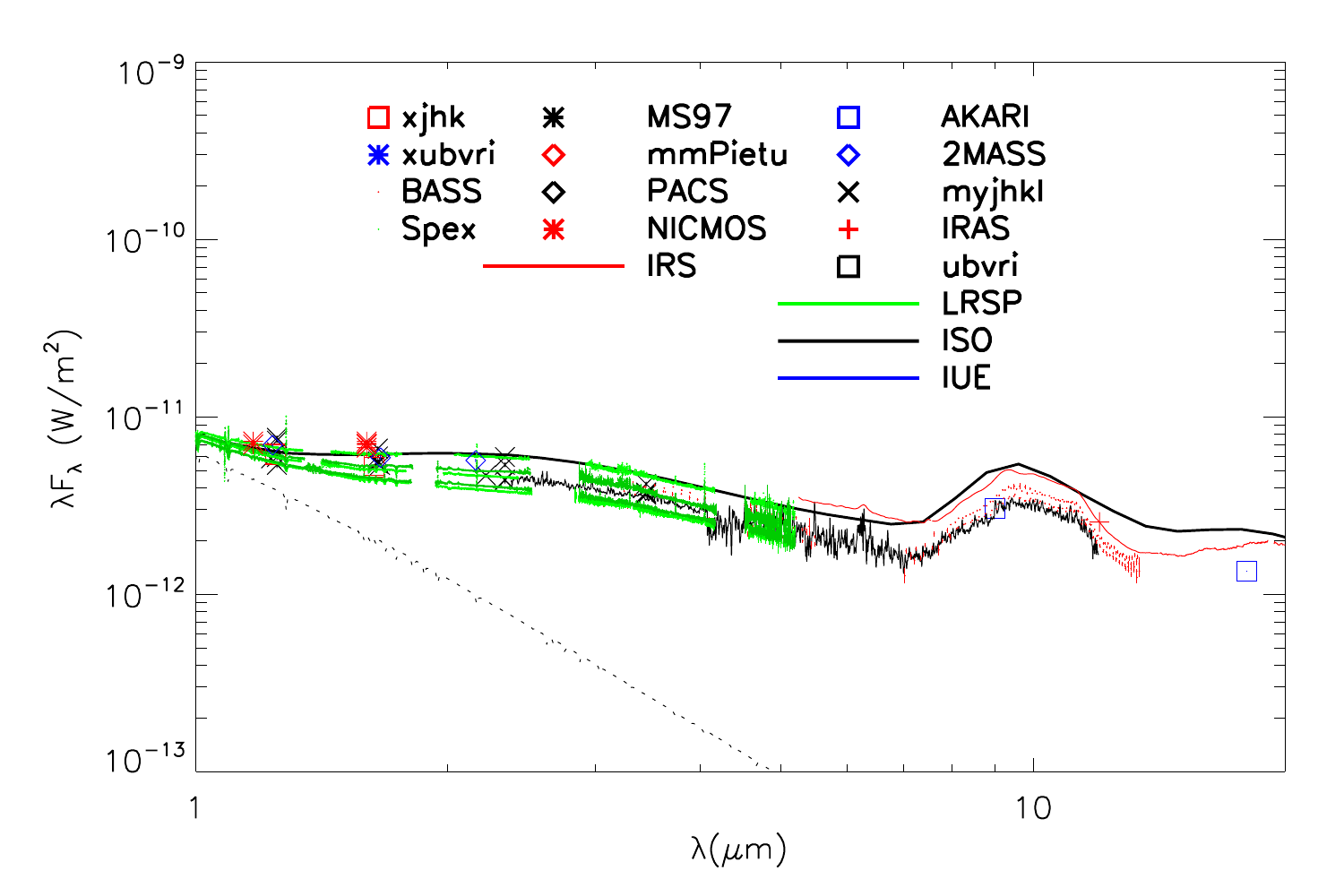} 
\caption{SED of the high state with the disk wind geometry. By changing the density of the wind component's dropoff with distance $r$ from $\rho \sim r^{-1.2}$ to $\rho \sim r^{-1.8}$, and pushing the actual inner envelope gap radius from 1 au to 2.5 au, the model produces far-IR emission that, in contrast to the model using a changing inner rim puff scale height, are still consistent with the data.}
\end{figure}

\begin{figure}[H]
\figurenum{13}
\centering
\plottwo{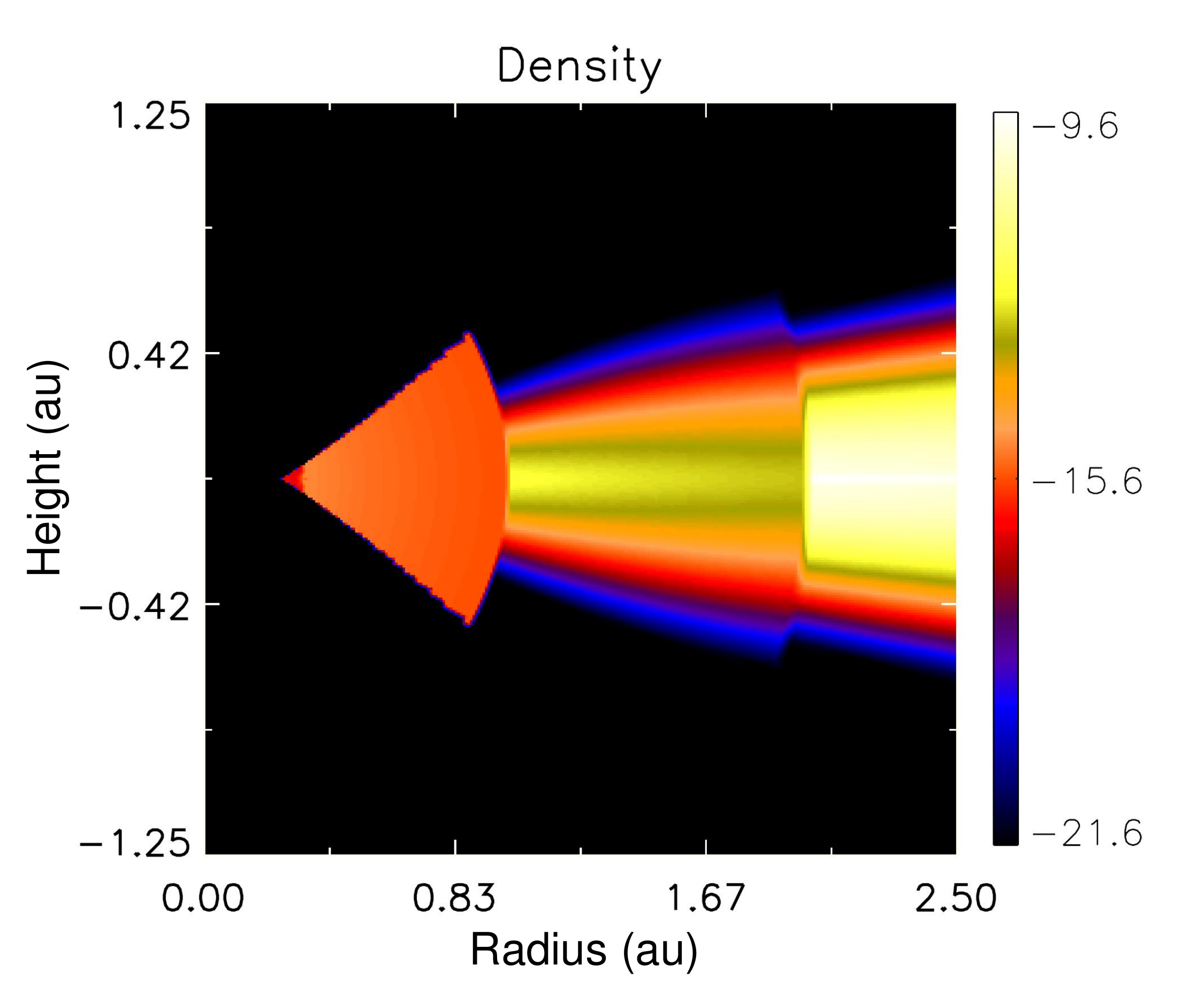}{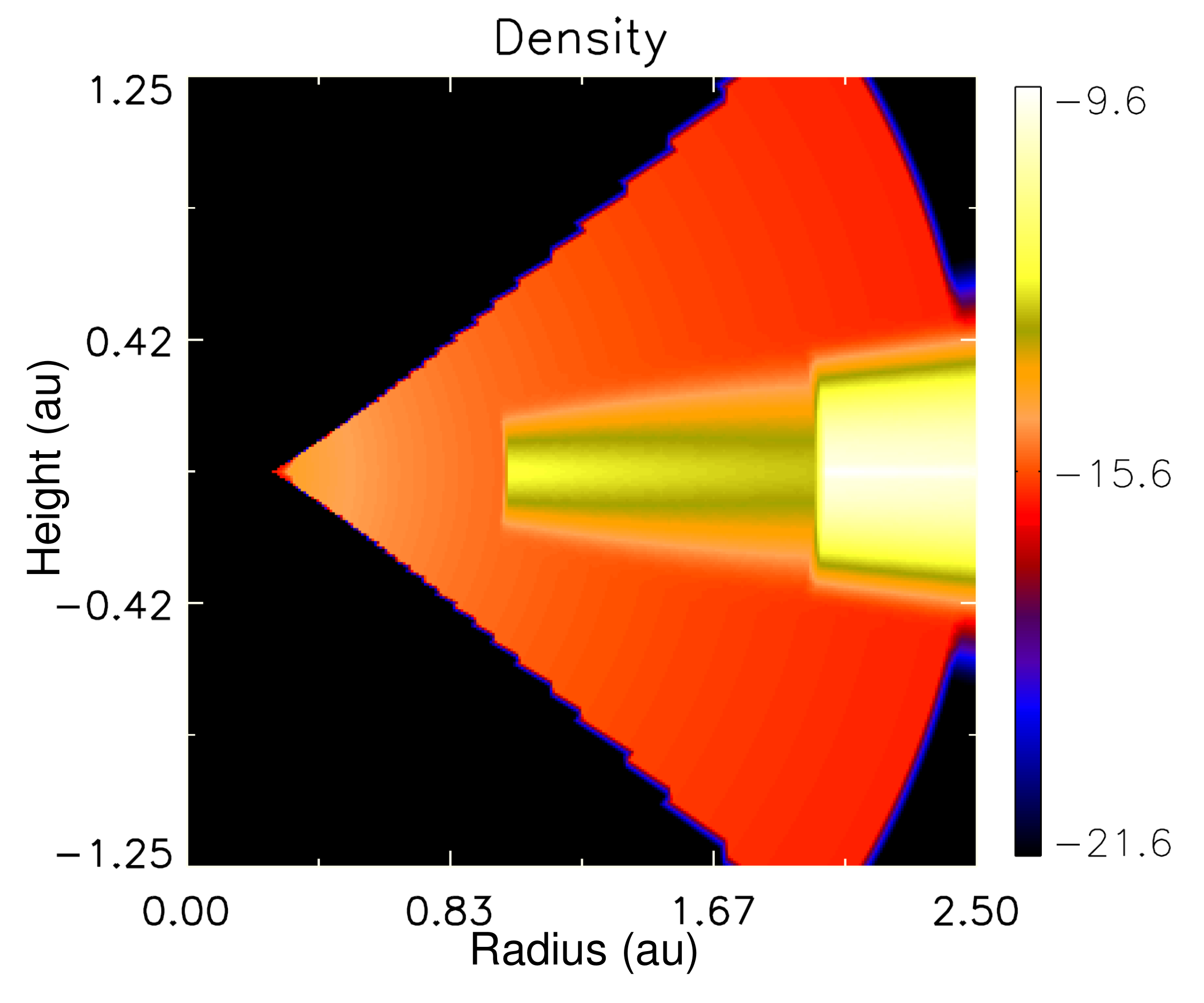}
\caption{A comparison of the inner disk region for the NIR low stater (left) and NIR high state (right) using a change in the inner jet structure instead of changing the puffed-up inner rim.}
\end{figure}

By contrast, in the variable jet/wind models, little change in the longer-wavelength SED occurred. This is more consistent with the ensemble of data at those wavelengths without the need to invoke the possible smearing of a variable response due to the time scale of the propagation of thermal response in the disk. In Figure 14, we calculated the surface  brightness of these ``jet'' models in the same manner as for the variable inner rim models. The minimal response to the light received \& scattered by the disk seems to be more  consistent with both the mid-IR data and lack of comparable response in the model SED at those wavelengths. \\

\begin{figure}[H]
\figurenum{14}
\centering
\includegraphics[scale=0.7]{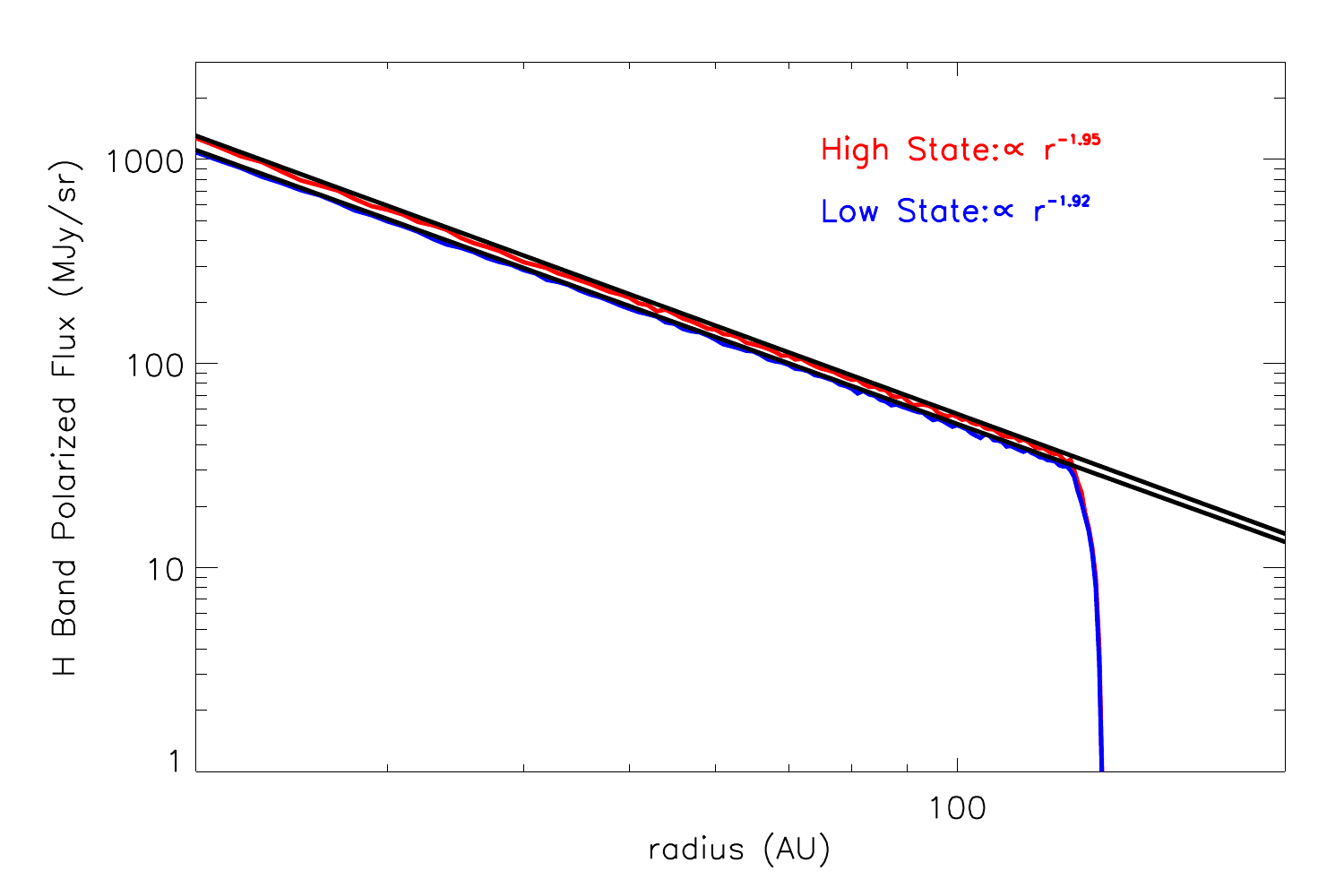}
\caption{The surface brightness for the two ``jet'' models calculated in the same manner as for the variable inner rim models. Here the change in the wind that fits the higher and lower near-IR observations produces little difference in the scattered light (as well as little difference on the mid-IR SED .}
\end{figure}

\section{Discussion}

We have investigated how changing the structure of the inner disk in MWC 480 could explain the observed variability seen at near-IR wavelengths. We find that changing the scale height of the inner rim of the dust disk, which has been successful in describing the variability in other objects, does not seem to apply in the case of MWC 480. Such changes should induce a drop in the flux at longer wavelengths (``see-saw'' behavior) which is inconsistent with existing data. And while the thermal response sue to changes in stellar illumination could potentially wash out the response at longer wavelengths, the presence of such a smoothed-out response in many disks systems indicates that this may not explain the ``stable'' fluxes seen at longer wavelengths, Instead, a more optically thin inner disk wind-like structure is able to produce changes in the NIR without producing far IR changes. Since it is known that MWC 480 possesses outflowing bipolar material, such a model would be required to be consistent with its presence. The principle difference between the models explored here is that the wind model trades optical depth for solid angle as seem by the star. The wind model is much more extended vertically, but has a lower overall mean optical depth than the inner disk rim description. It would be useful to search for the presence of jet-like structures in other object with near-IR variability where evidence for ``see-saw'' behavior is weak or absent.\\

\citet{Kusakabe12} detected the scattered light in the disk of MWC 480 at a time when the near-IR flux was at an historic minimum. This would seem to \textit{favor} the variable inner rim models. However, this could be due to a couple of factors. The first is that the historic low state at the time of the HiCIAO observations is even lower that those we are modeling, based primarily on the SpeX and BASS observations. It would be important to obtain multi-epoch scattered light images, as has been done for HD 163296 (e.g. \citep{Grady10,monnier17}). Likewise, multi-epoch interferometry in the near-IR would be important to obtain. \citep{Tannirkulam08a,Tannirkulam08b} detected changes in the inner regions of HD 163296. However, comprehensive temporal coverage of MWC 480 is lacking. \citet{jamialahmadi17} present 3 epochs of VLTI and Keck Interferometer data using 2-telescope beam combiners, for a total of 6 points in the uv plane. \citet{Lazareff17} have a single epoch using the 4-telescope PIONIER instrument on the VLTI. A more comprehensive study will be required to obtain a better understanding of the structural changes in the inner regions of MWC 480.

\section{Conclusion}

While changes in the scale height of the inner rim of young stellar disks has been implicated in their observed near-IR variability, such structural changes predict measurable variations in the disk flux at wavelengths dominated by more distant regions of the disk - the mid- to far-IR. Despite the $\sim$30\% changes observed in the near-IR, flux of MWC 480, there exists little evidence that the corresponding changes at longer wavelengths is occurring. Rather, changes in a more vertically extended, but optically thinner, structure seems capable of explaining the large variability in the near-IR, but without a similar change at longer wavelengths. Such a structure is already known to exist in MWC 480, as evidenced by the bipolar jets seen in deep imaging, and should be included in any model. Such structures have also been implicated in the variability in other similar stars, such as HD 163296. It is ironic that these two objects, that were part of the ``inspiration'' for the development of the variable inner rim scenario \citep{Sitko08}, should be better-explained by a different mechanism - changes in a wind-like structure - that are also consistent with the observed inference (we do not see their base directly) of the existence of such structures. \\

A more rigorous test of these models for MWC 480 can be made with truly contemporaneous NIR and MIR observations. Such an investigation could be done with a combination of ground-based NIR observations, with MIR data obtained with facilities such as NASA's Wide-Field Infrared Survey Telescope - Astrophysics-Focused Telescope Assets (WFIRST-AFTA) or James Webb Space Telescope (JWST). Despite the relative brightness brightness of the star, coronagraphic imaging of the disk will be possible with JWST. \\

MLS is supported by NASA Exoplanet Research Program grants NNX16AJ75G and NNX17AF88G. CAG is supported under NASA Origins of Solar Systems Funding  via NNG16PX39P. This research has made use of the NASA/ IPAC Infrared Science Archive, which is operated by the Jet Propulsion Laboratory, California Institute of Technology, under contract with the National Aeronautics and Space Administration. Some of the data presented in this paper were obtained from the Mikulski Archive for Space Telescopes (MAST). STScI is operated by the Association of Universities for Research in Astronomy, Inc., under NASA contract NAS5-26555. Support for MAST for non-HST data is provided by the NASA Office of Space Science via grant NNX09AF08G and by other grants and contracts. \\

The results reported herein benefitted from collaborations and/or information exchange within NASA's Nexus for Exoplanet System Science (NExSS) research coordination network sponsored by NASA's Science Mission Directorate. \\

\software{Spextool (Cushing et al. 2004), HOCHUNK3D (Whitney et al. 2013)}

\clearpage

\appendix

\section{Model Parameters}

\begin{table}
\begin{center}
\begin{tabular}{ |p{8cm}||p{4cm}|p{4cm}| }
 \hline
 \multicolumn{3}{|c|}{Stellar Parameters of MWC 480} \\
 \hline
 Parameter & Value &  Source\\
 \hline
 RA(J2000) & $04^{h} 58^{m} 46.271^{s}$ & (1)\\
 DEC(J2000) & $+29\degr 50' 376.61"$ & (1)\\
 Distance & 142$\pm$7 pc & (2)\\
 V & 7.62 & (3)\\
 B-V & 0.16 & (3)\\
 Luminosity & 15.1$\pm$1.5 L$_{\astrosun}$  & . . . \\
 Mass & 2.15$\pm$0.05 M$_{\astrosun}$ & (4)  \\
 Age & 7.5$\pm$ 1.5 Myr & (4)\\ 
 Disk Inclination & 37.5$\degr$ & (5)\\
 \hline
\end{tabular}
\end{center}
 \caption{1. Gaia Archive 2. Gaia Archive. Older Hipparcos value: 137$\pm$24 pc 3. \citep{mendigutia12} 4. Derived from Pisa PMS tracks \citep{tognelli11} 5. Rout from \citet{Kusakabe12}}
\end{table}

\begin{table}
\begin{center}
\begin{tabular}{ |p{8cm}||p{4cm}|p{4cm}|  }
 \hline
 \multicolumn{3}{|c|}{Model Parameters} \\
 \hline
 Parameter & Value & Source\\
 \hline
 Disk mass (total)& 0.07 M$_{\astrosun}$ & . . .\\ 
 Fraction of mass in large grains disk & 0.2 & . . .\\
 Settled disk grain file & www003.par & (6) \\
 Settled disk scale height normalization & 0.005 R$\star$ & . . .\\
 Rin-Rout & 0.15-200 AU & (5)\\  
 Settled disk radial density $r^{-A}$ &  A = 2.0 & . . .\\
 Settled disk scale height $r{^B}$ & B = 1.0 & . . .\\
 Envelope grain file & kmh & (7)\\
 Flared disk grain file & mrn77 & (8)\\
Non-settled disk scale height normalization & 0.028 R$\star$ & . ..\\
Non-settled Rin-Rout & 0.15-200AU & (5)\\
Non-settled radial density $r^{-A}$ & A = 2.15 & . . .\\
Non-settledscale height $r^B$ & B = 1.15 & . . .\\
 \hline
\end{tabular}
\end{center}
\caption{6. Model 1 from \citet{wood02} 7. \citet{kmh94} 8. \citet{mrn77}}
\end{table}

\clearpage

\section{Mass and Age of MWC 480}

Using the unweighted mean V magnitude of MWC 480 from \citet{Oudmaijer01}  (the range in V here is 0.06 mag based on 5 observations, that  in \citet{dewinter01} was 0.04 mag, based on 3 observations, while a range of 0.11 was derived from the Hipparcos Hp observations \citep{ESA97} that covers both B and V filter wavelengths), the distance from the first-release Gaia archive, and the bolometric correction and effective temperature for an A3 PMS stars from \citet{pecaut13} we derived a bolometric luminosity of 15.1$\pm$1.5 L$_{\astrosun}$ and effective temperature of 8550 K. Figure 9 shows the location of MWC relative to the PMS evolution tracks and isochrones of the non-gray models of \citet{tognelli11}, for z=0.02, helium mass fraction of 0.27, convective parameter $\alpha$ = 1.68, and a deuterium abundance of 2 x 10$^{-5}$. The on-line versions of these tracks and isochrones use the abundances of \citet{asplund05} but are nearly indistinguishable from those based on the revised abundances of \citet{asplund09} as seen in the figures in Tognelli et al. From these we derived a mass of 2.15$\pm$0.05 M$_{\astrosun}$ and an age of 7.5$\pm$1.5 Myr.

\begin{figure}[H]
\figurenum{14}
\centering
\includegraphics[scale=0.7]{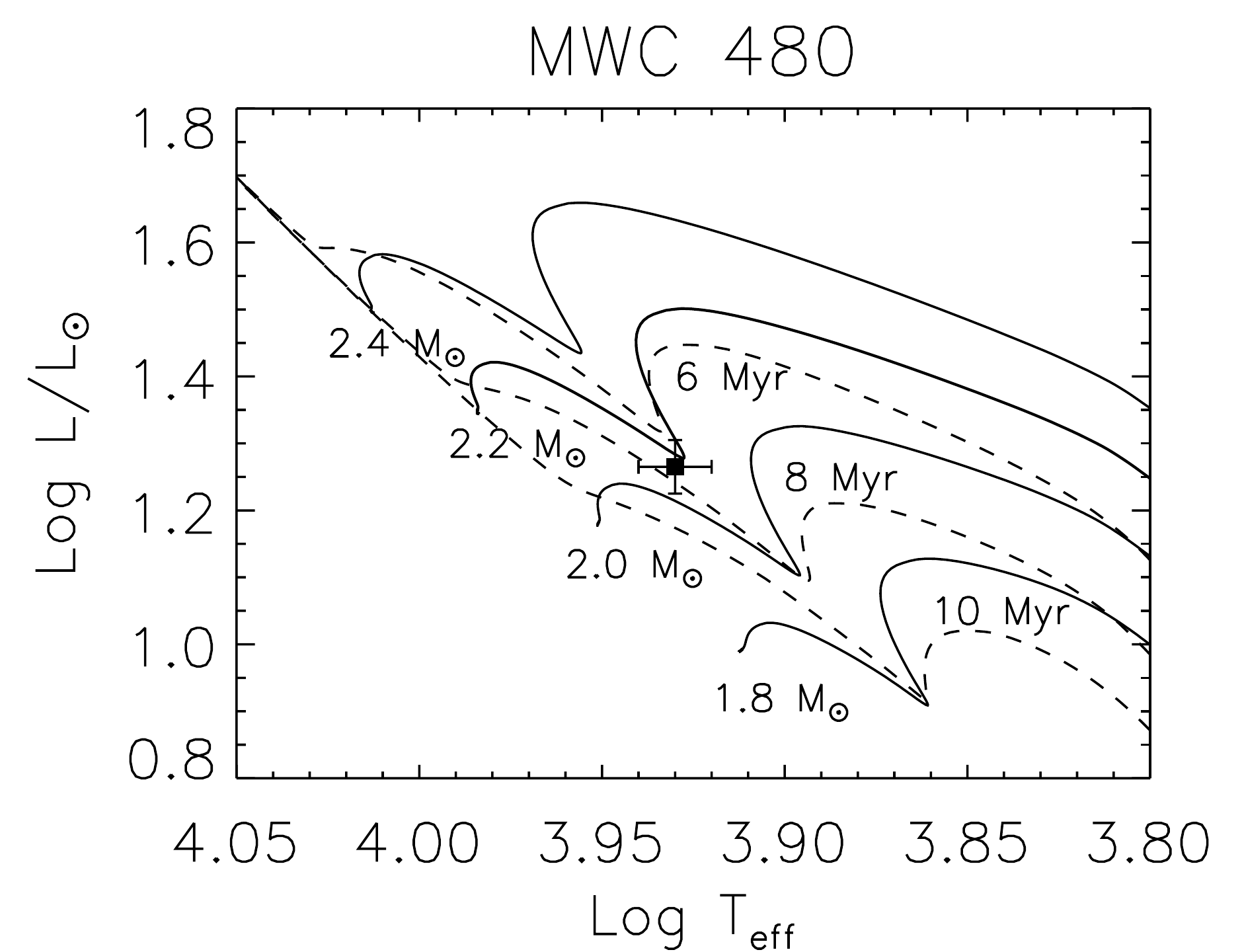}
\caption{The location of MWC 480 relative to the PMS evolutionary tracks and isochrones of \citet{tognelli11}.}
\end{figure}

\end{document}